\documentclass[12pt]{article}
\usepackage{epsf}
\usepackage{rotating}
\usepackage{amssymb}
\usepackage{amsmath}
\catcode`\@=11
\@addtoreset{equation}{section}
\def\theequation{\arabic{equation}}
\def\theequation{\thesection\arabic{equation}}
\newcommand{\appendixA}{\setcounter{equation}{0}
\def\theequation{\rm{A}.\arabic{equation}}\section*}

\catcode`\@=11

\def\section{\@startsection{section}{1}{\z@}{3.5ex plus 1ex minus
   .2ex}{2.3ex plus .2ex}{\large\bf}}
\def\thesection{\arabic{section}.}

\def\ps@headings{\def\@oddfoot{}\def\@evenfoot{}
\def\@oddhead{\hbox{}\hfill
        \makebox[.5\textwidth]{\raggedright\ignorespaces --\thepage{}--
        \hfill }}
\def\@evenhead{\@oddhead}
\def\subsectionmark##1{\markboth{##1}{}} }
\newcommand{\PLB}[3]{\emph{ Phys.~Lett.} \textbf{B#1} (#2) #3}

\def\dalemb#1#2{{\vbox{\hrule height .#2pt
        \hbox{\vrule width.#2pt height#1pt \kern#1pt
                \vrule width.#2pt}
        \hrule height.#2pt}}}

 \let\b=\beta

  \let\u=\upsilon

\def\nn{\nonumber} \def\bd{\begin{document}} \def\ed{\end{document}}
\def\ds{\documentstyle} \let\fr=\frac \let\bl=\bigl \let\br=\bigr
\let\Br=\Bigr \let\Bl=\Bigl
\let\bm=\bibitem
\let\na=\nabla
\let\pa=\partial \let\ov=\overline
\def\ie{{\it i.e.\ }}
\newcommand{\pr}{\paragraph{}}
\newcommand{\be}{\begin{equation}}
\newcommand{\ee}{\end{equation}}
\newcommand{\beba}{\begin{equation}\begin{array}{lcl}}
\newcommand{\eaee}{\end{array}\end{equation}}
\newcommand{\td}{\tilde}
\newcommand{\norsl}{\normalsize\sl}
\newcommand{\ns}{\normalsize}
\newcommand{\refs}[1]{(\ref{#1})}
\def\simlt{\mathrel{\lower2.5pt\vbox{\lineskip=0pt\baselineskip=0pt
           \hbox{$<$}\hbox{$\sim$}}}}
\def\simgt{\mathrel{\lower2.5pt\vbox{\lineskip=0pt\baselineskip=0pt
           \hbox{$>$}\hbox{$\sim$}}}}
\newcommand{\Mvariable}[1]{#1}
\renewcommand{\arraystretch}{1.1}
\newcommand{\bms}[5]{$({\bf#1},{\bf#2},#3,#4,#5)$}
\newcommand{\sms}[3]{$({\bf#1},{\bf#2},#3)$}
\newcommand{\ct}[1]{\tilde{c_{#1}}}
\renewcommand{\b}[1]{\textbf{#1}}
\newcommand{\bb}[1]{\bar{\textbf{#1}}}
\newcommand{\ba}{\begin{eqnarray}}
\newcommand{\ea}{\end{eqnarray}}
\renewcommand{\u}[1]{{U(#1)}}
\newcommand{\su}[1]{{SU(#1)}}
\usepackage{epsfig}
\newcommand{\ord}[2]{{\cal O}\left(#1\right)^{#2}}
\def\lsim{\mathrel{\rlap{\lower4pt\hbox{\hskip1pt$\sim$}}
    \raise1pt\hbox{$<$}}}                % less than or approx. symbol
\def\gsim{\mathrel{\rlap{\lower4pt\hbox{\hskip1pt$\sim$}}
    \raise1pt\hbox{$>$}}}                % greater than or approx. symbol
\begin{document}
\thispagestyle{empty} \enlargethispage{\baselineskip}
\rightline{\normalsize\sf hep-th/0210263} \rightline{\normalsize
CERN-TH/2002-272 }
 \rightline{\normalsize October 2002} \vskip
0.5truecm \centerline{\Large\bf D-branes and the Standard Model}
\vskip .5truecm
\centerline{{\large\bf I. Antoniadis}~$^a$\footnote{On leave of absence from
CPHT, UMR du CNRS 7644, Ecole Polytechnique, 91128 Palaiseau,
France},
{\large\bf E. Kiritsis~$^{b,d}$},
{\large\bf J. Rizos~$^c$}  and {\large\bf T.N. Tomaras}~$^{b}$}
\vskip .3truecm
\centerline{{\it $^a$CERN Theory Division
CH-1211, Gen{\`e}ve 23, Switzerland}}
%\vskip .5truecm
\centerline{\it $^b$Department of Physics and Institute of Plasma Physics, University of
Crete}
\centerline{\it and FO.R.T.H., 710 03 Heraklion, Greece}

\centerline{\it $^c$Department of Physics, University of Ioannina, 45110
Ioannina, Greece}
\centerline{\it $^d$CPHT, UMR du CNRS 7644, Ecole Polytechnique, 91128 Palaiseau,
France}
\vskip 0.5truecm
\centerline{\bf\small ABSTRACT}
\vskip .3truecm
We perform a systematic study of the Standard Model embedding in a D-brane
configuration of type I string theory at the TeV scale. We end up with an
attractive model and we study several phenomenological questions,
such as gauge coupling unification, proton stability, fermion masses and neutrino
oscillations. At the string scale, the gauge group is
$U(3)_{color}\times U(2)_{weak}\times U(1)_1\times U(1)_{bulk}$.
The corresponding gauge bosons are localized on three collections of branes; two
of them describe the strong and weak interactions, while the last abelian factor
lives on a brane which is extended in two large extra dimensions with a size of a
few microns. The hypercharge is a linear combination of the first three $U(1)$s.
All remaining $U(1)$s get masses at the TeV scale due to
anomalies, leaving the baryon and lepton numbers as (perturbatively) unbroken
global symmetries at low energies. The conservation of baryon number assures
proton stability, while lepton number symmetry guarantees light neutrino masses
that involve a right-handed neutrino in the bulk.
The model predicts the value of the weak angle which is compatible with the
experiment when the string scale is in the TeV region. It also contains
two Higgs doublets that provide tree-level masses to all fermions of the heaviest
generation, with calculable Yukawa couplings; one obtains a naturally heavy
top and the correct ratio $m_b/m_\tau$.
We also study neutrino masses and mixings in relation to recent
solar and atmospheric neutrino data.

%\hfill\break
%\vfill\eject

\section{Introduction}

In a previous work \cite{Antoniadis:2000en,akt2}, a minimal embedding of
the Standard Model (SM) was proposed in a D-brane configuration of type I
string theory with large internal dimensions and low fundamental scale
\cite{ Antoniadis:1998ig, Antoniadis:1990ew}. The $SU(3)$ color
and $SU(2)$ weak gauge fields were confined on two different collections of
branes. The model correctly accommodated the right value of the weak angle for a
choice of the string scale of a few TeV. It contained two Higgs doublets and
guaranteed proton stability. Among the issues, which were not addressed, are the
fermion masses, neutrino oscillations, and a natural suppression of lepton number
violating processes.

A generic feature of the models studied was that some
of the SM states should correspond to open strings with one end in the
bulk, implying the existence of some extra branes, in addition to the ones used
above \cite{Antoniadis:2000en,akt2}.
Starting from the last point, in the present work we introduce an extra brane in
the bulk with a corresponding $U(1)_b$ bulk gauge group \cite{akt2}. This group is broken by
anomalies, leaving behind an additional global symmetry that will be identified
with the lepton number.
In order to give masses to the neutrinos, we introduce a right-handed neutrino in
the bulk \cite{bulknu} that carries non-trivial lepton number. Large neutrino
masses are then forbidden by symmetry, while the right-neutrino coupling
suppression required to explain the neutrino oscillation data, is achieved if the
bulk has two dimensions of submillimeter size.

More precisely, in the minimal case of one bulk neutrino,
we show that solar and atmospheric neutrino data
can be accommodated using essentially the two lowest frequencies
of the neutrino mass matrix: the mass of the zero mode, arising
via the electroweak Higgs phenomenon, which is suppressed by the
volume of the bulk, and the mass of the first Kaluza--Klein (KK)
excitation. The former is used to reproduce the large mixing angle
(LMA or even LOW) solution to the solar neutrino anomaly, through
$\nu_e\leftrightarrow\nu_\mu$ transitions. The later is used to
explain atmospheric neutrino oscillations with an amplitude which
is enhanced due to logarithmic corrections of the two-dimensional
bulk \cite{AB}. Compatibility of the two conditions using one bulk
right neutrino is possible only if one introduces a non-orthogonal
angle between the two compact bulk dimensions, that leads
simultaneously to a CP violation in the neutrino sector.
Atmospheric oscillations contain however a significant sterile
component which seems to be in contradiction with recent atmospheric data analyses.
% which will be soon experimentally tested.

We also compute the tree-level Yukawa couplings of the two higgses to the
fermions of the heaviest generation. They are given in terms of the gauge
couplings and lead to a naturally heavy top and a ratio $m_b/m_\tau$ compatible
with the experimental data. Next, we proceed to a systematic description of the
main features that we will use in the following sections.

The general framework is type I string theory. We shall restrict
ourselves to models in which the closed string sector is
supersymmetric, while supersymmetry is generically broken by the
open strings at the string scale \cite{bsb}. \footnote{Recent
progress in constructing type I vacua with structure close to the
SM can be found in
\cite{Cremades:2002dh,obl,typeiims}.}
Within our framework, the minimal ensemble of D-branes needed in
our construction is the following mutually orthogonal stacks: a
stack of three coincident branes to generate the color group, a
second stack of two coincident branes to describe the weak
$SU(2)_L$ gauge bosons, and one more brane to generate the
$U(1)_b$ bulk discussed above. The resulting gauge group so far is
$U(3)_c\times U(2)_L\times U(1)_b$, with the three $U(1)$
generators denoted by $Q_c$, $Q_L$ and $Q_b$, respectively. To
ensure proton stability, we require baryon number conservation
with generator $B\equiv Q_c$. The hypercharge $Y$ cannot have a
component along $Q_b$, since this would lead to unrealistically
small gauge coupling, and as explained in \cite{Antoniadis:2000en}
the correct assignment of SM quantum numbers requires the presence
of an extra abelian factor, named $U(1)_1$ with generator $Q_1$,
living on an additional brane. This brane should lie on top of the
color or the weak stack of branes, as we argue below.

Since in our framework, supersymmetry is broken by combinations of (anti)branes
and orientifolds which preserve different subsets of the bulk supesymmetries, any
pair of D-branes D$p$ and D$p'$ satisfy $p-p'=0$ mod 4. It follows that a system
with three stacks of mutually orthogonal branes in the six-dimensional internal
(compact) space consists, up to T-dualities, of D9-branes with two different types
of D5-branes, extended in different directions. Specifically, the $U(1)_b$ lives
on the D9-brane, while the $U(3)_c$ and $U(2)_L$ are confined on two stacks of
5-branes, the first along say the 012345 and the other along the 012367 directions
of ten-dimensional spacetime. Thus, the (submillimeter) bulk is necessarily
two-dimensional (extended along the 89 directions), and the additional $U(1)_1$
brane has to coincide with either $U(3)_c$ or $U(2)_L$. The parameters of the
model are the string scale $M_s$, the string coupling $g_s$ and the volumes
$v_{45}$,
$v_{67}$ and $v_{89}$ of the corresponding subspaces, in string
units.\footnote{Using T-duality, we choose all internal volumes to be bigger than
unity, $v_{ij}>1$.} In terms of those, the four-dimensional Planck mass $M_P$ is
given by
\ba
M_P^2=\frac{8}{g_s^2} v_{45}v_{67}v_{89} M_s^2
\label{mp2}
\ea
and the non-abelian gauge couplings are
\ba
\frac{1}{g_3^2}=\frac{1}{g_s} v_{45}\qquad ;\qquad
\frac{1}{g_2^2}=\frac{1}{g_s} v_{67}
\label{g32}
\ea
It follows that
\ba
M_P^2=\frac{8}{g_3^2 g_2^2}v_{89}M_s^2=
\frac{2}{\alpha_3\alpha_2}{\hat v}_{89}M_s^2 \, ,
\label{mp}
\ea
where $\alpha_i=g_i^2/4\pi$ and ${\hat v}_{89}\equiv v_{89}/(2\pi)^2=R_8R_9$ for a
rectangular torus of radii $R_8, R_9$. The $U(1)_1$ gauge coupling $g_1$ is
equal to $g_3$ ($g_2$), if the $U(1)_1$ brane is on top of the $U(3)_c$
($U(2)_L$).

Upon T-duality, one finds two
additional realizations: (i) a set of D3-branes (along 0123) describing $U(3)_c$,
and two orthogonal sets of D7-branes along 01236789 and 01234567 describing
$U(1)_b$ and $U(2)_L$, respectively; (ii) three sets of D5-branes along 012389,
012345 and 012367, giving rise to $U(1)_b$, $U(3)_c$ and $U(2)_L$,
respectively. In both cases, relation (\ref{mp}) remains intact.

The gauge coupling $g_b$ of the $U(1)_b$ gauge boson which lives in the bulk is
extremely small since it is suppressed by the volume of the bulk $v_{89}$. For
instance, in the case where the $U(1)_b$ lives on a D9-brane, its coupling is
given by
\ba
\frac{1}{g_b^2}=\frac{1}{g_s}v_{45}v_{67}v_{89}=\frac{g_s}{8}\frac{M_P^2}{M_s^2}\, ,
\label{gb}
\ea
where in the second equality we used eq.~(\ref{mp2}). Using
now the weak coupling condition $g_s<1$ and the inequality
$g_s>g_{3,2}^2$ following from $v_{ij}>1$ in eq.~(\ref{g32}), one finds
\ba
{\sqrt 8}\frac{M_s}{M_P}<g_b<\frac{{\sqrt 8}}{g_3}\frac{M_s}{M_P}\, ,
\label{gbin}
\ea
which implies that $g_b\simeq 10^{-16} - 10^{-14}$ for $M_s\sim 1-10$ TeV.

If the $U(1)_b$ gauge boson is light, it will be subject to strong contraints
coming from supernova observations, since it would be copiously produced in
various nuclear reactions leading to supernova cooling through energy loss in the
bulk of extra dimensions. The corresponding process is much stronger than the
production of gravitons because of the non-derivative coupling of the
gauge boson interaction \cite{add2}. In fact, in the case of $n$ large transverse
dimensions of common radius $R$, satisfying $m_A,R^{-1}<<T$ with $m_A$ the gauge
boson mass and $T$ the supernova temperature, the production rate $P_A$ is
proportional to
\ba
P_A\sim g_b^2\times [R(T-m_A)]^n\times\frac{1}{T^2}\simeq\frac{T^{n-2}}{M_s^n}\, ,
\ea
where the factor $[R(T-m_A)]^n$ counts the number of Kaluza--Klein
(KK) excitations of the $U(1)_b$ gauge boson with mass less than $T$. This rate can be
compared with the corresponding graviton production
\ba
P_G\sim\frac{1}{M_P^2}\times (RT)^n\simeq\frac{T^n}{M_s^{n+2}}\, ,
\ea
showing that for $n=2$ (sub)millimeter extra dimensions, it is unacceptably large,
unless the bulk gauge boson acquires a mass $m_A\simgt 10$ MeV.

The paper is organized in seven sections, of which this introduction is the first.
In Section 2, we perform a systematic search for models with four sets of
branes corresponding to the gauge group $U(3)_c\times U(2)_L\times
U(1)_1\times U(1)_b$ with the minimal standard model fermion spectrum and
a Higgs sector that generates masses for all quarks and leptons of the
heaviest generation. We identify the hypercharge $U(1)_Y$
combination and in Section 3 we
perform a renormalization group analysis of gauge
couplings to identify models with low string scale, where the $U(1)_1$ is on
top of either the color or the weak branes.
In Section 4, we select four models with string scale in the TeV region,
possessing in addition baryon and lepton number conservation, and we describe
their main phenomenological features.\footnote{Orientifold models with baryon
and lepton number conservation were also constructed in Ref.~\cite{obl}.} They
all contain two Higgs doublets that can provide tree-level masses to all
fermions. Moreover, apart from the hypercharge, all other
abelian factors are broken by
mixed gauge and gravitational anomalies and become massive at the string
scale. In Section 5, we compute the tree-level Yukawa couplings of the two
higgses to the fermions of the heaviest generation and study predictions for
mass relations. In Section 6, we introduce one right-handed neutrino in the
bulk and study the generation of neutrino masses and neutrino oscillations.
%. We show that the resulting
%be made compatible with both the solar and atmospheric
%neutrino data with the .
Finally, Section 7 contains our summary and conclusions.

\section{Model search}

As shown in \cite{Antoniadis:2000en}, the minimal $D$-brane configuration
that can successfully accommodate the Standard Model (SM)
consists of three sets of branes with gauge
symmetry $\u3_c\times\u2_L\times\u1_1$. The first set contains three
coincident branes (``color" branes). An open string with one end
attached to this set transforms as an $\su3_c$ triplet (or anti-triplet),
but also carries an additional $\u1_c$ quantum number which
can be identified with the (gauged) baryon number.
Similarly, $\u2_L$ is realized by a set of two coincident branes
(``weak" branes) and open strings attached to them from the one end are $SU(2)_L$
doublets characterized by an additional
$\u1_L$ quantum number, the (gauged) weak ``doublet" number. Moreover,
consistency of the SM embedding requires
the presence of an additional $\u1_1$ factor, generated by a single
brane. This is needed for several reasons: TeV scale unification,
baryon number conservation, and mass generation for all quarks and leptons of the
heaviest generation. The hypercharge is then a linear combination of the three
abelian factors,
$Y=k_3\,Q_c+k_2\,Q_L+k_1\,Q_1$,
where $Q_c,Q_L,Q_1$ are the
charges under $\u1_c,\u1_L,\u1_1$ respectively. It turns out
\cite{Antoniadis:2000en}
that there exist four possible ``viable" models that
reproduce the weak mixing angle al low energies.
They correspond to $k_3=\frac{2}{3}$ ($k_3=-\frac{1}{3}$), $k_2=\pm\frac{1}{2}$,
$k_1=1$ and require the abelian brane $\u1_1$ to be on top of the
color (weak) branes, so that $g_3=g_1$ ($g_2=g_1$).

In all the above brane configurations there exist states (e.g. the $SU(2)_L$
singlet anti-quarks) which correspond to open strings with only one of their ends
attached to one of the three sets of D-branes. The other end is in the bulk,
and requires the existence of some additional branes extended in the bulk,
carrying extra quantum numbers.
In this work, we consider a minimal extension of the models considered in
\cite{Antoniadis:2000en} by introducing one additional D-brane in the bulk
giving rise to an extra abelian gauge factor $\u1_b$. As we will see later, the
requirement of baryon and lepton number conservation leads to four possible models
that we are going to study in the next section. However, in this section, we
do not impose this constraint and we systematically explore the possibility of
reproducing the SM spectrum, together with possibly additional Higgs scalars, as
open strings stretched between any two of the four sets of branes. The extension
of the Higgs sector is required for the realization of the electroweak symmetry
breaking and mass generation for all fermions of at least one (the heaviest)
generation.

Thus, the total gauge group is
\ba
G &=& U(3)_c\times{U(2)}_L\times{U(1)}_1\times{U(1)}_b\nonumber\\ &=&
SU(3)_c\times{U(1)_c}\times{SU(2)}_L\times{{U(1)}_L}\times{U(1)}_1\times{U(1)}_b
\ea
and contains four abelian factors. The assignment of the SM particles is
partially fixed from its non-abelian structure. The quark doublet $Q$ corresponds
to an open string with one end on the color and the other on the weak set of
branes. The anti-quarks $u^c, d^c$ must have one of their ends attached to the
color branes. The lepton doublet and possible Higgs doublets must have one end on
the weak branes. However, there is a freedom related to the abelian structure,
since the hypercharge can arise as a linear combination of all four abelian
factors. In a generic model, the abelian charges can be expressed without loss of
generality in terms of ten parameters displayed in Table~\ref{tta}.
\begin{table}[!ht]
\center
\begin{tabular}{|c|c|c|c|c|c|c|}
\hline
particle&${U(1)}_c$&${U(1)}_L$&${U(1)}_1$&${U(1)}_b$\\
\hline
$Q(\b3,\b2,\hphantom{+}\frac{1}{6})$&$+1$&$w$&$0$&$0$\\
$u^c(\bb3,\b1,-\frac{2}{3})$&$-1$&$0$&$a_1$&$a_2$\\
$d^c(\bb3,\b1,+\frac{1}{3})$&$-1$&$0$&$b_1$&$b_2$\\
$L\;(\b1,\b2,-\frac{1}{2})$&$0$&$+1$&$c_1$&$c_2$\\
$e^c(\b1,\b1,+1)$&$0$&$d_L$&$d_1$&$d_2$\\
\hline
\end{tabular}
\caption{\label{tta} SM particles with their generic charges under
 the abelian part of
the gauge group $U(3)_c\times{U(2)}_L\times{U(1)}_1\times{U(1)}_b$.}
\end{table}

In a convenient parametrization, normalizing the $U(N)\sim SU(N)\times U(1)$
generators as ${\rm Tr} T^a T^b=\delta^{ab}/2$, and
measuring the corresponding $U(1)$ charges with respect to the coupling
$g/\sqrt{2N}$, the ten parameters are integers:
$a_{1,2},b_{1,2},c_{1,2},d_2=0,\pm1$, $d_{1}=0,\pm1,\pm2$, $d_L=0,\pm2,w=\pm1$
satisfying
\ba
\sum_{i=1,2}|a_i|=\sum_{i=1,2}|b_i|=\sum_{i=1,2}|c_i|=1,\
\sum_{i=1,2,L}|d_i|=2\, .  \label{range}\
\ea
The first three constraints in (\ref{range}) correspond to the
requirement that the $u^c$ and $d^c$ anti-quarks, as well as the lepton doublet,
must come from open strings with one end attached to one of the abelian
D-brane sets. The fourth constraint forces the positron $e^c$ open string
to be stretched either between the two abelian branes, or to have both ends
attached to the abelian $\u1_1$ brane, or to the weak set of branes. In the
latter case, it has $U(1)_L$ charge $\pm 2$ and is an $SU(2)_L$ singlet
arising from the antisymmetric product of two doublets.
The parameter $w$ in Table~\ref{tta}
refers to the $\u1_L$ charges of the quark-doublets, that we can choose to
be $\pm1$, since doublets are equivalent with anti-doublets.
Note that a priori one might also consider the case in which one of the $u^c$
and $d^c$ anti-quarks arises as a string with both ends on the
color branes $({\bf 3}\times{\bf 3}={\bf \bar 3}+{\bf 6})$, so that its
$\u1_c$ charges would be $\pm2$. This, however, would invalidate the identification of
$\u1_c$ with the baryon number and forbid the presence of quark mass terms, since
one of the combinations $Q u^c$ and $Q d^c$ would not be neutral under
$\u1_c$. Hence, this case will not be explored.

The hypercharge can in general be a linear combination
of all four abelian group factors. However, we restrict ourselves to
models in which the bulk ${U(1)}_b$ does not contribute to the
hypercharge, in order to avoid an unrealistically small gauge coupling.
Hence,
\ba
Y= k_3\,Q_c+k_2\,Q_L+k_1\,Q_1\, .\label{ourh}
\ea
The correct assignments for SM particles are reproduced, provided
\ba
k_3+k_2\,w=\frac{1}{6}\nonumber\\
-k_3+a_1 k_1=-\frac{2}{3}\nonumber\\
-k_3+b_1\,k_1=\frac{1}{3}\label{hyp}\\
k_2+c_1\,k_1=-\frac{1}{2}\nonumber\\
k_2\,d_L+d_1\,k_1=1.\nonumber
\ea
Notice that the second and third of the above equations imply that $k_1\ne0$.

The next step, after assigning the correct hypercharge to the SM particles,
is to check for the existence of  candidate fermion mass terms.
Here, we discuss only the question of masses for one generation (the heaviest) and
we do not address the general problem of flavor. To lowest order, the mass terms
are of the form $Q d^c H^\dagger_d$, $Q u^c H_u$ and
$L e^c H_e^\dagger$ where $H_d, H_u, H_e$ are scalar Higgs doublets
with appropriate charges. In a generic model,
there are four different candidate Higgs
scalar doublets (and their conjugates) $H_1,\dots,H_4$, with
$\u1_L\times\u1_1\times\u1_b$ charges:
\ba
\left\{H_1,H_2,H_3,H_4\right\}=\left\{(1,1,0),(1,0,1),(1,-1,0),(1,0,-1)\right\}
\, .
\label{Hcharges}
\ea
It is easy to show that for
any hypercharge embedding of the form (\ref{ourh}) with $k_1\ne0$, there are at most
three of the above Higgs doublets that have the correct hypercharge. Depending on the
parameters of the model, they can be reduced to two.
For the generic charge assignments of Table~\ref{tta}, the required Higgs
charges are
\ba
H_u&=&\left(\b1,\b2,0,-w,-a_1,-a_2\right)\nonumber\\
H_d&=&\left(\b1,\b2,0,+w,+b_1,+b_2\right)\label{allh}\\
H_e&=&\left(\b1,\b2,0,1+d_L,c_1+d_1,c_2+d_2\right)\, .
\nonumber
\ea

Provided the constraints (\ref{range}) are satisfied, both $H_u$
and $H_d$ have the right charges of (\ref{Hcharges}) and correspond to strings
stretched between the weak and one of the abelian branes. Thus, (\ref{range})
guarantees the existence of tree-level quark masses. On the other hand, the
existence of $H_e$ depends on the particular choice of parameters, e.g. for
$c_1+d_1=2$, $H_e$ does not exist
and a tree-level lepton mass term ($L e^c H^\dagger$)
is forbidden. The generic constraint that guarantees tree-level
lepton masses is
\ba
\sum_{i=1,2}\left|c_i+d_i\right|=
\left|1+d_L\right|=1\ .\label{masset}
\ea
Given the smallness of the lepton mass compared to the masses
of the quarks, of the same generation, it would be reasonable
to examine also the possibility that the lepton mass is generated
by a higher order term. The next order
candidate lepton mass term is of dimension six, proportional to
$\frac{1}{M^2_s}\,L e^c H^\dagger H^\dagger H$.
The constraint in this case is more complicated and the method we
are going to use is the following: for each configuration that
satisfies all other constraints except (\ref{masset}), we
derive explicitly the candidate Higgs doublets and check the
existence of possible fifth order mass terms.\footnote{Here, we check only the
conservation of all gauge quantum numbers. In the string context, there may be
additional selection rules for the non-vanishing of the corresponding couplings
that are model dependent.}

The hypercharge constraints  (\ref{hyp}) can be easily solved.
They require $a_1\ne b_1$ and
\ba
k_3&=&\frac{{a_1} + 2\,{b_1}}
  {3\,\left( {b_1} -{a_1} \right) }\\
k_2&=&-\frac{\left( {a_1} + {b_1} \right)
      }{2\,\left(  {b_1}-{a_1}
      \right)\, w }\\
k_1&=&\frac{1}{b_1-a_1}\\
c_1&=&-\frac{{b_1} - {a_1}}{2} +
  \frac{\left( {a_1} + {b_1}
       \right) }{2w}\\
d_1&=& {b_1}-{a_1}  +
  \frac{\left( {a_1} + {b_1}
       \right) \,{d_L}}{2w}\, .
\ea
The allowed values of $(a_1,b_1)$ are
$\left\{(-1,0),(-1,1),(0,-1),(0,1),(1,-1),(1,0)\right\}$. However, we notice that
the solutions with parameters $(a_1,b_1,c_1,d_1, k_1)$ and
$(-a_1,-b_1,-c_1,-d_1,-k_1)$ are equivalent, since they correspond to a global
change of sign $Q_1\to -Q_1$. Thus, it is sufficient to search for solutions with
$(a_1,b_1) \in \left\{(-1,0),(-1,1),(0,1)\right\}$.
Solving for these choices, we get three allowed hypercharge embeddings:
\ba
(i) \ \ \ a_1=-1, b_1=1 &:& Y=\frac{1}{6} Q_c +\frac{1}{2} Q_1 \label{emba}\\
(ii) \ \ a_1=-1, b_1=0 &:& Y=-\frac{1}{3} Q_c +\frac{w}{2} Q_L+  Q_1 \label{embb}\\
(iii) \ \ \ a_1=0, b_1=1 &:& Y=\frac{2}{3} Q_c -\frac{w}{2} Q_L+  Q_1\, .
\label{embc}
\ea
Case (i)  leads to
$c_1=-1, c_2=0, d_1=2, d_1=d_L=0$. This is a special solution where the $\u1_b$ brane
decouples from the model since no SM particles are attached to it.  It satisfies
(\ref{masset}) and thus leads to tree level lepton masses. The solution exists
for both $w=\pm1$, as the value of
$w$ does not play an important role when $k_2=0$.
In case (ii), we have $c_1=-(1+w)/2$,
$d_L=0, d_1=1$ or $c_1=(1+w)/2, d_L=2w, d_1=d_2=0$, while
case (iii) leads to
$c_1=(w-1)/2, d_L=0, d_1=1$ or $c_1=(1+w)/2, d_L=2w, d_1=d_2=0$.

Combining the above three cases with the constraints (\ref{range}) and
(\ref{masset}), we get 9 distinct configurations with tree-level quark and lepton
masses, displayed in the upper part of Table~\ref{ptab}. Relaxing the
constraint (\ref{masset}) with the requirement that lepton masses arise
through dimension six effective operators, leads to 6 more distinct models
corresponding to the cases 10-15 of Table~\ref{ptab}. In deriving these
configurations, we have eliminated all models connected to the ones above by
the global charge redefinition $Q_b\to -Q_b$.

\begin{sidewaystable}
\begin{tabular}{|r|c|c|c|c|c|c|c|c|c|c|c|c|c|c|}
\hline
&$a_1$&$a_2$&$b_1$&$b_2$&$c_1$&$c_2$&$d_1$&$d_2$&$d_L$&$w$&$Y$&$L$&$n_h$\\
\hline
$1$&$-1$&$0$&$1$&$0$&$-1$&$0$&$2$&$0$&$0$&$1$&
$ \frac{1}{6}\,Q_c +  \frac{1}{2}\,Q_1$&$-$&$2$\\
$2$&$-1$&$0$&$0$&$-1$&$0$&$-1$&$1$&$1$&$0$&$-1$&
$-  \frac{1}{3}\,Q_c -  \frac{1}{2}\,Q_L+Q_1$
&$\hphantom{+}\frac{1}{2}\,Q_c +
\frac{1}{2}\,Q_L-\frac{1}{2}Q_1-\frac{1}{2}\,Q_b$&$2$\\
$3$&$-1$&$0$&$0$&$-1$&$-1$&$0$&$1$&$1$&$0$&$1$&
$-  \frac{1}{3}\,Q_c + \frac{1}{2}\,Q_L+Q_1$&$-$&$3$\\
$4$&$0$&$1$&$1$&$0$&$0$&$-1$&$1$&$1$&$0$&$1$&
$\hphantom{+}\frac{2}{3}\,Q_c-  \frac{1}{2}\,Q_L+Q_1$
&$-\frac{1}{2}\,Q_c +  \frac{1}{2}\,Q_L-\frac{1}{2}Q_1-\frac{1}{2}\,Q_b$
&$2$\\
$5$&$0$&$1$&$1$&$0$&$-1$&$0$&$1$&$1$&$0$&$-1$&
$\hphantom{+}\frac{2}{3}\,Q_c+\frac{1}{2}\,Q_L+Q_1$&$-$&$3$\\
$6$&$-1$&$0$&$0$&$-1$&$0$&$-1$&$0$&$0$&$-2$&$-1$&
$-  \frac{1}{3}\,Q_c-  \frac{1}{2}\,Q_L+Q_1$
&$\hphantom{+}\frac{1}{2}\,Q_c +
\frac{1}{2}\,Q_L-\frac{1}{2}Q_1-\frac{1}{2}\,Q_b$&$2$\\
$7$&$-1$&$0$&$0$&$1$&$0$&$-1$&$0$&$0$&$-2$&$-1$&
$-  \frac{1}{3}\,Q_c -  \frac{1}{2}\,Q_L+Q_1$&$-$&$3$\\
$8$&$0$&$1$&$1$&$0$&$0$&$1$&$0$&$0$&$-2$&$\hphantom{+}1$&
$\hphantom{+}\frac{2}{3}\,Q_c-  \frac{1}{2}\,Q_L   +Q_1$&$-$&$3$\\
$9$&$0$&$1$&$1$&$0$&$0$&$-1$&$0$&$0$&$-2$&$\hphantom{+}1$&
$\hphantom{+}\frac{2}{3}\,Q_c-  \frac{1}{2}\,Q_L + Q_1$
&$-\frac{1}{2}\,Q_c +  \frac{1}{2}\,Q_L-\frac{1}{2}Q_1-\frac{1}{2}\,Q_b$&$2$\\
\hline
$10$&$-1$&$0$&$0$&$1$&$-1$&$0$&$0$&$0$&$2$&$\hphantom{+}1$&$-  \frac{1}{3}\,Q_c+\frac{1}{2}\,Q_L+Q_1$&$-$&$3$\\
$11$&$0$&$1$&$1$&$0$&$-1$&$0$&$0$&$0$&$2$&$-1$&$\hphantom{+}\frac{2}{3}\,Q_c+\frac{1}{2}\,Q_L+Q_1$&$-$&$3$\\
$12$&$-1$&$0$&$0$&$1$&$0$&$1$&$1$&$1$&$0$&$-1$&$-  \frac{1}{3}\,Q_c -  \frac{1}{2}\,Q_L+Q_1$&$-$&$3$\\
$13$&$-1$&$0$&$0$&$-1$&$0$&$1$&$1$&$1$&$0$&$-1$&$-  \frac{1}{3}\,Q_c -  \frac{1}{2}\,Q_L+Q_1$&$-$&$3$\\
$14$&$0$&$1$&$1$&$0$&$0$&$1$&$1$&$1$&$0$&$\hphantom{+}1$&$\hphantom{+}\frac{2}{3}\,Q_c-  \frac{1}{2}\,Q_L+Q_1$&$-$&$3$\\
$15$&$0$&$1$&$1$&$0$&$0$&$-1$&$1$&$-1$&$0$&$\hphantom{+}1$&$\hphantom{+}\frac{2}{3}\,Q_c-  \frac{1}{2}Q_L+Q_1$&$-$&$3$\\
\hline
\end{tabular}
\caption{\label{ptab}\it Distinct models with lepton masses generated either at
tree level $L e^c H^\dagger$ (cases 1-9), or by dimension six effective
operators $\sim L e^c H^\dagger \langle H^\dagger H \rangle/M^2$ (cases
10-15). We also display the lepton number combination $L$ (when it exists) and
the number of Higgs doublets $n_h$, needed to generate quark and lepton
masses.}
\end{sidewaystable}

As we mentioned before, in all the above configurations, we can define the baryon
number $B$ as
\ba
B=\frac{1}{3} Q_c\, .
\ea
As we will argue below, $U(1)_c$ gauge invariance is broken by anomalies to a global
symmetry, implying baryon number conservation in type I string perturbation theory.
Since lepton number is also conserved at present energies, we can
further examine which of the above models possess also the lepton number $L$ as a
symmetry. In general, $L$ can also be expressed as a linear combination of
all abelian factors,
\ba
L=\sum_{i=c,L,1,b} p_i Q_i
\ea
 that satisfies
\ba
p_c+p_L w&=&0\nn\\
-p_c +a_1 p_1 +a_2 p_b&=&0\nn\\
-p_c +b_1 p_1 +b_2 p_b&=&0\label{lepdef}\\
 p_L +c_1 p_1 +c_2 p_b&=&1\nn\\
d_L p_L +d_1 p_1 +d_2 p_b&=&-1\nn
\ea
Inspection of (\ref{lepdef}), in conjunction with (\ref{hyp}) that requires
$a_1\ne b_1$, implies that lepton number can only be defined for $p_b\ne0$,
i.e. only in the presence of the bulk $U(1)_b$. This is of course expected,
since the models without $U(1)_b$ have no lepton number
\cite{Antoniadis:2000en}. Solving explicitly (\ref{lepdef}) for each one of the
cases of Table~\ref{ptab}, we find that only four models, namely 2,4,6,9,
incorporate the lepton number as a (gauged) abelian symmetry. Its precise
definition for each of these models is also presented in the table.

\section{The weak angle and the string scale}

We now come to the determination of the string scale consistent
with the low energy SM data.
Following the hypercharge definition (\ref{ourh}),
the low energy data depend on the couplings $g_3$, $g_2$ and $g_1$ of the three
brane sets $\u3_c$, $\u2_L$ and $\u1_1$. These couplings are in principle
independent, but, as already explained in the introduction, in order to lower the
string scale we have to consider configurations where the $\u1_1$ brane is on top
of either the $\u3_c$ or the $\u2_L$ stacks. Hence, we have two possible coupling
relations at the string scale
\ba
(i)\ \ g_3=g_1\qquad \mbox{or}\qquad (ii) \ \ g_2=g_1\ . \label{couprel}
\ea

In our normalizations, the hypercharge coupling
$g_Y$ at the string scale
is expressed as
\ba
\frac{1}{g_Y^2}=\frac{6\,k_3^2}{g_3^2}+\frac{4\,k_2^2}{g_2^2}+\frac{2\,k_1^2}{g_1^2}\, .
\label{yy}
\ea
Following the one loop coupling evolution ($\alpha_i={g_i^2}/{4\pi}$),
\ba
\frac{1}{\alpha_i(M_s)}=\frac{1}{\alpha_i(M_Z)}+\frac{b_i}{4 \pi}
\ln\frac{\Delta^I M_s}{M_Z}\, ,
\ea
where $b_3=-7, b_2=-10/3+n_h/6,
b_Y=20/3+n_h/6$ and  $n_h$ is the number of scalar Higgs doublets.
The constant $\Delta^I$ corresponds to a model independent piece of
the type I string thresholds, entering in the identification of the string scale
with the ultraviolet cutoff of the effective field theory. Its value was computed in
Ref. \cite{AQ} to be $\Delta^I=1/\sqrt{\pi e^\gamma}\simeq 0.4$, where $\gamma$
is the Euler's constant. When the string scale is very high compared to present
energies, this represents a small correction compared to the dominant
logarithmic contribution coming from the renormalization group evolution. On the
other hand, when the string scale is low, such a correction becomes important, as
it effectively changes the string scale by roughly a factor of two, and should be taken
with caution since it is of the same order with the (unknown) model dependent
part of threshold corrections. Consequently, we will leave $\Delta^I$ as a
parameter and discuss its possible effects on our results case by case.

Solving the one-loop renormalization group equations (RGE) for the coupling
evolution, the values of the weak mixing angle $\sin^2\theta_W$ and of the
strong coupling $a_3$ at the $Z$-mass $M_Z$ are related to the couplings
at the string scale:
\ba
\sin^2\theta_W(M_Z)&=&\frac{1}{1+k_Y}+\frac{\alpha_{em}(M_Z)}{2\pi}
\frac{(k_Y b_2-b_Y)}{(1+k_Y)}\ln\frac{\Delta^I M_s}{M_Z}+\label{sine}\\
&~&\frac{\alpha_{em}(M_Z)}{1+k_Y}\left[
6k_3^2\left(\frac{1}{\alpha_L(M_s)}-\frac{1}{\alpha_3(M_s)}\right)
+2k_1^2\left(\frac{1}{\alpha_L(M_s)}-\frac{1}{\alpha_1(M_s)}\right)
\right]\nonumber\\
\frac{1}{a_3(M_Z)}&=&\frac{1}{\alpha_{em}(M_Z)}\frac{1}{1+k_Y}-\frac{1}{2\pi}
\frac{b_2+b_Y-b_3(1+k_Y)}{1+k_Y}\log\frac{\Delta^I M_s}{M_Z}+\label{ase}\\
&~&\frac{1}{1+k_Y}\left[
(4 k_2^2+1)\left(\frac{1}{\alpha_L(M_s)}-\frac{1}{\alpha_3(M_s)}\right)
-2 k_1^2\left(\frac{1}{\alpha_L(M_s)}-\frac{1}{\alpha_1(M_s)}\right)
\right]\nonumber
\ea
where $k_Y=6\,k_3^2+4\,k_2^2+2 k_1^2$ and $\alpha_{em}$ is the electromagnetic coupling.

Given a coupling relation of (\ref{couprel}), we can use eqs.~(\ref{sine}) and
(\ref{ase}) to determine the string scale $M_s$ that correctly reproduces the low
energy data. Clearly, the solution depends on $|k_3|$, $|k_2|$ and $|k_1|$.
According to our previous analysis, there are three classes of models, which correspond
to the three possible hypercharge embeddings (\ref{emba}), (\ref{embb}) and
(\ref{embc}):
\ba
(i) &:& |k_3|=\frac{1}{6}\ ,\ |k_2|=0 \ , \ |k_1|=\frac{1}{2}\nn\\
(ii) &:& |k_3|=\frac{1}{3}\ ,\ |k_2|=\frac{1}{2} \ , \ |k_1|=1\nn\\
(iii) &:& |k_3|=\frac{2}{3}\ ,\ |k_2|=\frac{1}{2} \ , \ |k_1|=1
\label{kkk}
\ea
Using (\ref{sine}) and (\ref{ase}), for each of the embeddings
(\ref{kkk}) and the unification conditions (\ref{couprel}), we
computed the corresponding string ``unification" scale $M_U\equiv
\Delta^I M_s$. In our calculation we have used the following
values for the low energy quantities $a_3(M_Z)=0.119$,
$\sin^2\theta_W=0.231$, $a_{em}(M_Z)=1/127.934$. The results are
presented in Table~\ref{rt1}.

In the above calculations we have assumed that the number of
doublets $n_h$ is the minimum $n_h=2$ required by the model.
Of course, one can consider models with more doublets
which can be for instance replicas of these two. It would be thus
interesting to examine the dependence of the above results on the
number of doublets. To this end we can extract analytic formulas
regarding the unification scale $M_U$. For the case $g_1=g_3$, taking
for simplicity $k_1=1$ and $k_L=\pm\frac{1}{2}$, we find
\ba
\frac{3(4+7\,k_3^2)}{\pi}\,\log\frac{M_U}{M_Z}=\frac{1}{\alpha_{em}(M_Z)}
(1-2\,\sin^2\theta_W(M_Z))
-2(1+3 k_3^2)\frac{1}{\alpha_3(M_Z)}
\ea
which implies that at the one-loop $M_U$ is independent of the
number of doublets. Similarly, for
$g_1=g_2$ and $k_1=1,k_L=\pm\frac{1}{2}$, we have
\ba
\frac{50+126 k_3^2-n_h}{6
\pi}\,\log\frac{M_U}{M_Z}=\frac{1}{\alpha_{em}(M_Z)}(1-4\,\sin^2\theta_W(M_Z))
-6 k_3^2 \frac{1}{\alpha_3(M_Z)} \
\ea
where we find a very weak dependence. Obviously, the number of
doublets affects the value of the weak gauge coupling at $M_s$ and
thus the volume of the bulk through (\ref{mp}).

\begin{table}
\centering
\begin{tabular}{|r|c|c|c|c|c|c|c|}
\hline
&$|k_3|$&$|k_2|$&$|k_1|$&$M_U (TeV)$&$g_2(M_U)/g_3(M_U)$&$g_2(M_U)g_3(M_U)$\\
\hline
&$\frac{1}{6}$&$0$&$\frac{1}{2}$&$4.6\times 10^{20}$&$1.1$&$0.21$\\
\cline{2-7}
$g_1=g_3$&$\frac{1}{3}$&$\frac{1}{2}$&$1$&$2.4\times 10^3$&$0.76$&$0.48$\\
\cline{2-7}
&$\frac{2}{3}$&$\frac{1}{2}$&$1$&$7.2$&$0.65$&0.61\\
\hline
&$\frac{1}{6}$&$0$&$\frac{1}{2}$&$1.5\times 10^{22}$&$1.1$&$0.26$\\
\cline{2-7}
$g_1=g_2$&$\frac{1}{3}$&$\frac{1}{2}$&$1$&$0.32$&$0.57$&$0.73$\\
\cline{2-7}
&$\frac{2}{3}$&$\frac{1}{2}$&$1$&$-$&$-$&$-$\\
\hline
\end{tabular}
\caption{\it \label{rt1}
The string unification scale $M_U$ and the two
independent gauge couplings for the two possible brane configurations and the
various hypercharge embeddings.}
\end{table}

\section{The models}

So far, we have classified all possible
$\u3_c\times\u2_L\times\u1_1\times\u1_b$ brane models that can successfully
accommodate the SM spectrum. The quantum numbers of each model as well as the
hypercharge embedding are summarized in Table~\ref{ptab}. Furthermore,
compatibility with type I string theory with string scale in the TeV region,
requires the bulk to be two-dimensional of (sub)millimeter size, and leads to
two possible configurations: Place the $\u1_1$ brane on top of the weak
$\u2_L$ stack of branes or on top of the color $\u3_c$ branes. These impose
two different brane coupling relations at the string (unification) scale:
$g_1=g_2$ or $g_1=g_3$, respectively. For every model, using the hypercharge
embedding of Table~\ref{ptab}, the one loop gauge coupling evolution and one of
the above brane coupling conditions, we can determine the unification (string)
scale that reproduces the weak angle at low energies. The results are
summarized in Table~\ref{rt1}.

According to the results of Section 2, there are three distinct hypercharge embeddings
that correspond to $\left(|k_3|,|k_2|,|k_1|\right)$ $=$
$\left\{(1/6,0,1/2),(1/3,1/2,1),(2/3,1/2,1)\right\}$. Since we wish to restrict ourselves
to models in which supersymmetry is broken at the string scale ($M_s$), we
would like $M_s$ to be low, at the TeV scale, to protect the mass hierarchy. Thus,
model 1 of Table~\ref{ptab}, with hypercharge embedding $(1/6,0,1/2)$, is
rejected for both $\u1_1$ brane arrangements, since the resulting string scale
is too high. Furthermore, models with hypercharge embedding $(1/3,1/2,1)$ lead
to $M_s\sim 10^3$ TeV for $g_1=g_3$. This scale, although much lower than the
traditional GUT scale, is rather high for the  stabilization of  hierarchy. On
the contrary, for $g_1=g_2$, we find $M_s\sim{\cal O}(1)$ TeV (when the
universal threshold correction
$\Delta^I$ is taken into account) that lies at the edge of the present
experimental limits. The third embedding
$(2/3,1/2,1)$ reproduces successfully  the low energy data only for
$g_1=g_3$ and a string scale $M_s\sim{\cal O}(10)$ TeV.

In all configurations of Table~\ref{ptab}, the baryon number appears as a
gauged abelian symmetry. This symmetry is broken due to mixed gauge and
gravitational anomalies leaving behind a global symmetry. Baryon number
conservation is essential for low string scale models, since one needs to
eliminate effective operators to very high accuracy in order to avoid fast
proton decay, starting with dimension six operators of the form $Q Q Q L$
which are not sufficiently suppressed
\cite{Ibanez:1999it}.
%\footnote{For different proposals to suppress proton decay,
%see Refs.~\cite{protondecay}.}
In addition to baryon number, one should also assure that the
lepton number is a good symmetry of the low energy theory. Lepton number
conservation is also essential for preservation of acceptable neutrino masses, as
it forbids for instance the presence of the dimension 5 operator $L L H H$. Such an
operator would lead to large Majorana neutrino masses, of the order of a few GeV, in
models  where the string scale, typically a few TeV, is too low for the
operation of an effective sea-saw mechanism. Hence, we shall be interested only
in models in which the
lepton number is a good symmetry. Indeed, as seen in Table~\ref{ptab},
only in four models, namely 2,4,6 and 9, lepton number appears as a (gauged)
abelian symmetry. Being anomalous, this  symmetry will be broken, but
lepton number will survive as a global symmetry of the effective theory.

In fact, these four models can be derived in a straightforward way by simple
considerations of the quantum numbers. The quark doublet $Q$ is fixed by non abelian
gauge symmetries, while existence of baryon number implies that the antiquarks
$u^c, d^c$ correspond to strings stretched between the color branes and one each of the
abelian branes $U(1)_1$ and $U(1)_b$. Thus, one has two possibilities leading to
models that we call $A$ ($d^c$ has one end in the bulk) and $B$ ($u^c$ sees the
bulk). Existence of lepton number fixes the lepton doublet as a string stretched
between the weak branes and the $U(1)_b$ brane, while for each of the models
$A$ and $B$ there are two possibilities for the antilepton $e^c$ to emerge as a
string stretched between the two abelian branes, or to have both ends on the
weak branes. Thus, we obtain two additional models that we call $A'$ and $B'$.
As it can also be seen in the table, all these models have tree-level quark and
lepton masses and make use of only two Higgs doublets. They also require low energy
string scale for some of the brane coupling conditions. We now proceed to a
detailed study of these four models and to an analysis of their main
phenomenological characteristics.

Notice from Table~\ref{rt1} that in both classes of models $A$ and
$B$, the coupling constant ratio is $g_2/g_3\simeq 0.6$ at the
string scale, implying through the relations of Section 1 that at
least one of the internal compact dimensions along the
world-volume of the weak set of branes must be larger than the
string length, by at least a factor of two (in the case of two
large dimensions, or by a factor of four in the case of one). The
relevant experimental signal would be the production of
Kaluza--Klein excitations for the $W^\pm$ bosons and the other
mediators of the electroweak interactions but not of gluons,
providing one of the first indications of new physics \cite{KK}.

\medskip
\noindent{\it Models $A$ and  $A'$}

\begin{figure}
\center
\epsfxsize=6cm
\epsfbox{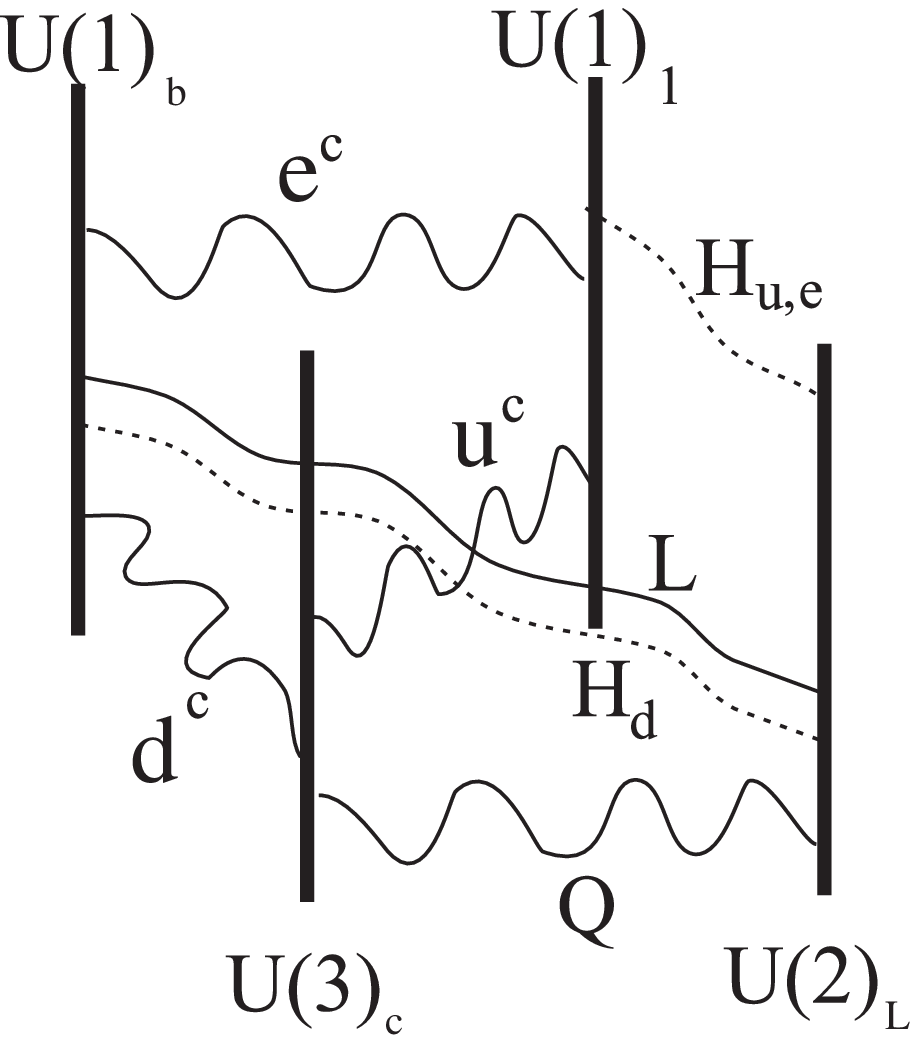}
\epsfxsize=6cm
\epsfbox{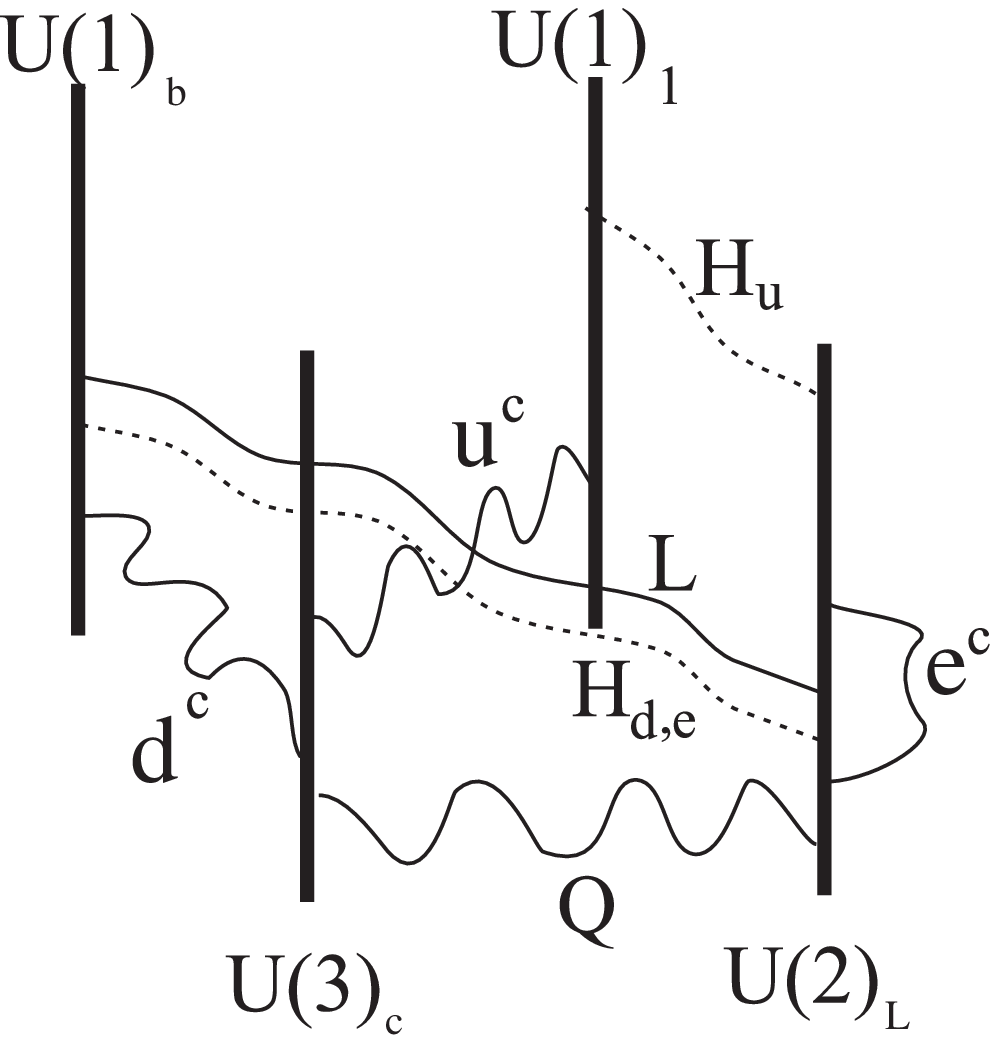}
\caption{\label{figa}{\it Pictorial representation of models} $A, A'$.}
\end{figure}

We consider here  the models 2 and 6 of Table~\ref{ptab},
hereafter referred as models $A$ and $A'$ respectively.
They are characterized by the common  hypercharge embedding
\ba
Y= -\frac{1}{3}\,Q_c-\frac{1}{2}\,Q_L+Q_1\label{hypa}
\ea
but they differ slightly in their spectra.
The spectrum of model $A$ is
%shown in the table\\
%\begin{tabular}{ccccccc}
%$Q$&$\b3$&$\b2$&$-1$&$+1$&$0$&$0$\\
%\end{tabular}
\ba
&~&Q\left(\b3,\b2,+1,-1,0,0\right)\nn\\
&~&u^c(\bb3,\b1,-1,0,-1,0)\nn\\
&~&d^c(\bb3,\b1,-1,0,0,-1)\nn\\
&~&L(\b1,\b2,0,+1,0,-1)\nn\\
&~&e^c(\b1,\b1,0,0,+1,+1)\nn\\
&~&H_u(\b1,\b2,0,+1,+1,0)\nn\\
&~&H_d(\b1,\b2,0,-1,0,-1)\nn
\ea
while in model $A'$ the right-handed electron $e^c$ is replaced by an open string
with both ends on the weak brane stack, and thus $e^c=(\b1,\b1,0,-2,0,0)$. The two
models are presented pictorially in Figure~\ref{figa}.

Apart from the hypercharge combination (\ref{hypa}) all remaining abelian
factors are anomalous. Indeed, for every abelian generator $Q_I, I=(c,L,1,b)$, we can
calculate the mixed gauge anomaly $K_{IJ}\equiv {\rm Tr} Q_I T^2_J$ with
$J=SU(3),SU(2),Y$, and gravitational anomaly $K_{I4}\equiv{\rm Tr} Q_I$
for both models $A$ and $A'$:
\ba
K^{(A)}=
\begin{pmatrix}
  0 & -1 & -\frac{1}{2} & -\frac{1}{2} \\
  \frac{3}{2} & -1 & 0 & -\frac{1}{2} \\
  -\frac{3}{2} & \frac{1}{3} & -\frac{1}{3} & \frac{1}{6} \\
  0 & -4 & -2 & -4
\end{pmatrix}\ ,\
K^{(A')}=
\begin{pmatrix}
  0 & -1 & -\frac{1}{2} & -\frac{1}{2} \\
  \frac{3}{2} & -1 & 0 & -\frac{1}{2} \\
  -\frac{3}{2} & -\frac{5}{3} & -\frac{4}{3} & -\frac{5}{6} \\
  0 & -6 & -3 & -5
\end{pmatrix}\label{anoma}
\ea It is easy to check that the matrices $K K^T$ for both models
have only one zero eigenvalue corresponding to the hypercharge
combination (\ref{hypa}) and three non vanishing ones
corresponding to the orthogonal $U(1)$ anomalous combinations. In
the context of type I string theory, these anomalies are cancelled
by a generalized Green-Schwarz mechanism which makes use of three
axions that are shifted under the corresponding $U(1)$ anomalous
gauge transformations \cite{gGS}. As a result, the three extra
gauge bosons become massive, leaving behind the corresponding
global symmetries unbroken in perturbation theory
\cite{Poppitz:1998dj}. The three extra $U(1)$'s can be expressed
in terms of known SM symmetries: \ba
\mbox{Baryon number}\,\ \ \ B&=&\frac{1}{3} Q_c\nn\\
\mbox{Lepton number}\,\,\ \ \ L&=&\frac{1}{2}\left(Q_c+Q_L-Q_1-Q_b\right)\label{glob}\\
\mbox{Peccei--Quinn}\ \ \ Q_{PQ}&=&-\frac{1}{2}\left(Q_c-Q_L-3\,Q_1-3\,Q_b\right)\nn
\ea
Thus, our effective SM inherits baryon and lepton number
as well as Peccei--Quinn (PQ) global symmetries from the anomaly cancellation
mechanism. Note however that $PQ$ is the original Peccei--Quinn symmetry only in
model $A'$, such that all fermions have charges $+1$, while $H_u$ and $H_d$ have
charges $-2$ and $+2$, respectively. In model $A$, the global $PQ$
symmetry defined in (\ref{glob}) is similar but with lepton charge
+3. The reason is that in model $A$ the fermion-Higgs Yukawa couplings are different,
and leptons get masses from $H_u$ and not from $H_d$.

The general one-loop string computation of the masses of anomalous $U(1)$ gauge
bosons, as well as their localization properties in the internal compactified
space, was performed recently for generic orientifold vacua \cite{AKR}.
It was shown that orbifold sectors preserving $N=1$ supersymmetry yield
four-dimensional (4d) contributions, localized in the whole six-dimensional (6d)
internal space, while $N=2$ supersymmetric sectors give 6d contributions
localized only in four internal dimensions. The later are related to 6d
anomalies. Thus, even $U(1)$s which are apparently anomaly free
may acquire non-zero masses at the one-loop level, as a consequence of 6d
anomalies \cite{AKR,SST}. These results have the following implications in our case:
\begin{enumerate}
\item
The two $U(1)$ combinations, orthogonal to the hypercharge and localized on the
strong and weak D-brane sets, acquire in general masses of the order of the
string scale from contributions of $N=1$ sectors, in agreement with effective
field theory expectations based on 4d anomalies.
\item
Such contributions are not sufficient though to make heavy the third $U(1)$
propagating in the bulk, since the resulting mass terms are localized and
suppressed by the volume of the bulk. In order to give string scale mass, one
needs instead $N=2$ contributions associated to 6d anomalies along the two large
bulk directions.
%====================================added Jan 03
In our models such contributions are indeed in general present
and arise from mixed 6d gauge-gravitational anomalies of two different sources:
(i) the generic presence of a neutral 6d Weyl fermion on the bulk brane which
coincides either with the $U(1)_b$ gaugino (in the supersymmetric case)
or the goldstino of the non-linearly realised
supersymmetry (in the brane SUSY breaking case \cite{bsb});
(ii) the contribution of the right-handed neutrino which arises from a
six-dimensional Weyl spinor. As a result, the third abelian gauge field $U(1)_b$
acquires also a mass of the order of the string scale, although its gauge coupling
is tiny due to the volume suppression (see eq. (\ref{gbin})).
%===================================added Jan 03
\item
Special care is needed to guarantee that the hypercharge remains massless despite
the fact that it is anomaly free, along the lines of Ref.~\cite{AKR}.
\end{enumerate}
%===================================added Jan 03
The presence of massive gauge bosons associated to anomalous abelian gauge symmetries
is generic. Their mass is given by $M_A^2\sim g_s M_s^2$,
up to a numerical model dependent factor and is somewhat smaller that the string scale.
When the latter is low, they can affect low energy measurable data, such as $g-2$
for leptons \cite{pana} and the $\rho$-parameter \cite{r}, leading to additional bounds
on the string scale.

Note that the global PQ symmetry leftover from $U(1)_b$ is spontaneously broken by
the Higgs expectation value giving rise to an unwanted electroweak axion. A possible
way out was suggested in ref. \cite{Antoniadis:2000en}, using an appropriate departure away from the
orientifold point.
%===================================added Jan 03

A plausible extension of the model is the introduction of a right-handed neutrino
in the bulk. A natural candidate state would be an open string ending on the
${U(1)}_b$ brane. Its charge is then fixed to $+2$ by the requirement of
existence of the single possible neutrino mass term $L\,H_d\,\nu_R$. The
suppression of the brane-bulk couplings due to the wave function of $\nu_R$ would
thus provide a natural explanation for the smallness of neutrino masses. Note
that if the zero mode of this bulk neutrino state is chiral, the anomaly
structure of the model changes: $B-L$ becomes anomaly free and as a consequence
the associated gauge boson remains in principle massless. However, as we
discussed above, this is not in general true because of 6d anomalies \cite{AKR}.
 In any case,
this problem is absent if we introduce a vector-like bulk neutrino pair
\ba
&~&{\nu}_R(\b1,\b1,0,0,0,+2)+{\nu}^c_R(\b1,\b1,0,0,0,-2)\nn
\ea
that leaves the anomalies (\ref{anoma}) intact. Note that ${\nu}^c_R$ does
not play any role in the subsequent discussion of neutrino masses and oscillations.
%===================================added Jan 03

Coming to the issue of gauge couplings and the string scale, as already
explained we have two different realizations for each model.
The first is with $g_1=g_3$ at $M_s$ that corresponds to a
configuration where the $\u1_1$ brane is placed on top of
the color branes. According to Table~\ref{rt1}, this leads to
an intermediate string scale $M_s\sim 10^6$ GeV, which appears too high to
guarantee the stabilization of hierarchy. The second possibility is to take the
$\u1_1$ brane on top of the weak branes, leading to $g_1=g_2$. The required
string scale is now low $M_s\sim{\cal O}(500)$ GeV (300-800 GeV, depending on the
threshold corrections), and could account for the stability of the hierarchy.

\medskip
\noindent{\it Models $B$ and $B'$}

Another  phenomenologically promising pair of models
consists of solutions 4 and 9  of Table~\ref{ptab}, named hereafter $B$ and
$B'$, which corresponds to the hypercharge embedding
\ba
Y= \frac{2}{3}\,Q_c-\frac{1}{2}\,Q_L+Q_1\, .
\label{hypb}
\ea
The spectrum is
\ba
&~&Q(\b3,\b2,+1,+1,0,0)\nn\\
&~&u^c(\bb3,\b1,-1,0,0,1)\nn\\
&~&d^c(\bb3,\b1,-1,0,1,0)\nn\\
&~&L(\b1,\b2,0,+1,0,-1)\nn\\
&~&e^c(\b1,\b1,0,0,+1,+1)\nn\\
&~&H_u(\b1,\b2,0,-1,0,-1)\nn\\
&~&H_d(\b1,\b2,0,+1,+1,0)\nn
\ea
for model $B$, while in $B'$ $e^c$ is replaced by $e^c(\b1,\b1,0,-2,0,0)$.
\begin{figure}
\center
\epsfxsize=6cm
\epsfbox{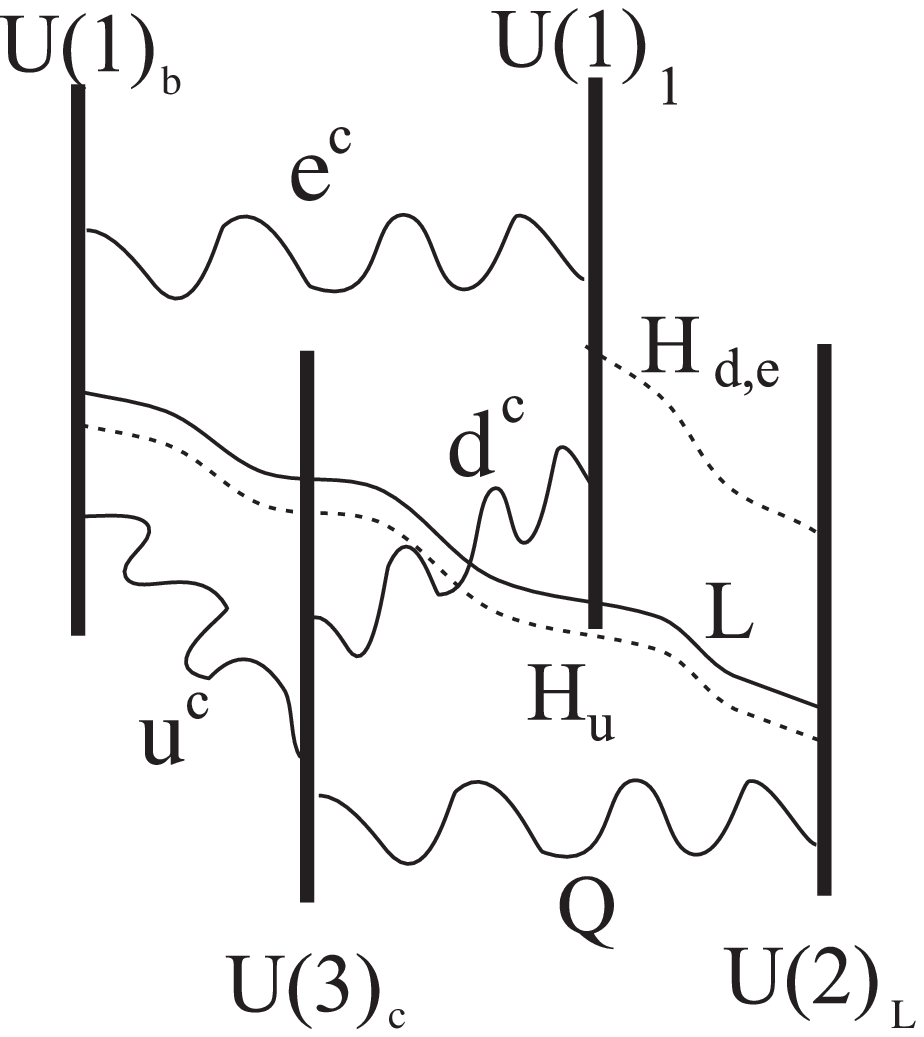}
\epsfxsize=6cm
\epsfbox{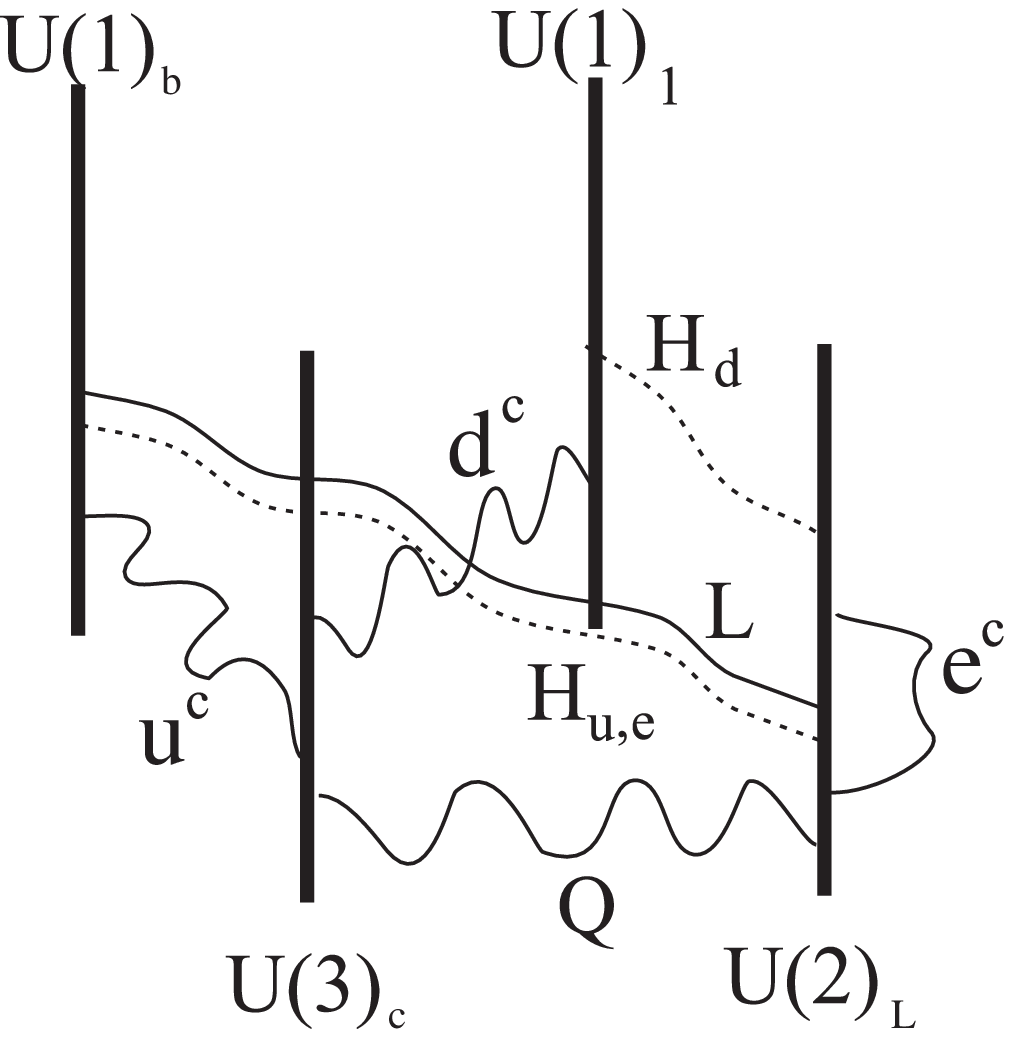}
\caption{\label{figb}\it Pictorial representation of models $B$ and $B'$.}
\end{figure}
The two models are represented pictorially  in Figure~\ref{figb}.
The four abelian gauge factors are anomalous. Proceeding as in the analysis
(\ref{anoma}) of models $A$ and $A'$, the mixed gauge and
gravitational anomalies are
\ba
K^{(B)}=
\begin{pmatrix}
  0 & 1 & \frac{1}{2} & \frac{1}{2} \\
  \frac{3}{2} & 2 & 0 & -\frac{1}{2} \\
  -\frac{3}{2} & \frac{2}{3} & \frac{4}{3} & \frac{11}{6} \\
  0 & 8 & 4 & 2
\end{pmatrix}\ ,\
K^{(B')}=
\begin{pmatrix}
  0 & 1 & \frac{1}{2} & \frac{1}{2} \\
  \frac{3}{2} & 2 & 0 & -\frac{1}{2} \\
  -\frac{3}{2} & -\frac{4}{3} & \frac{1}{3} & \frac{5}{6} \\
  0 & 6 & 3 & 1
\end{pmatrix}\label{anomb}
\ea
It is easy to see that the only anomaly free combination is the hypercharge (\ref{hypb})
which survives at low energies.
All other abelian gauge factors are anomalous and will be broken by the generalized
Green-Schwarz anomaly cancellation mechanism, leaving behind global symmetries. They can be
expressed in terms of the usual SM global symmetries as the following $\u1$
combinations:
\ba
\mbox{Baryon number}\,\ \ \ B&=&\frac{1}{3}\,Q_c\\
\mbox{Lepton number}\ \ \ \ L&=&-\frac{1}{2}\left(Q_c-Q_L+Q_1+Q_b\right)\\
\mbox{Peccei-Quinn}\ \ \
Q_{PQ}&=&\frac{1}{2}\left(-Q_c+3\,Q_L+Q_1+Q_b\right)
\ea
Similarly to the analysis of models $A$ and $A'$, the $PQ$ charges
defined above are the traditional ones only  for model $B$. In
model $B'$, the lepton charge is $-3$, as a result of the Higgs Yukawa
couplings to the fermions (see below).
The right handed neutrino can also be accommodated as an open string
with both ends on the bulk abelian brane:
\ba
&~&{\nu_R}(\b1,\b1,0,0,0,+2)+{\nu^c_R}(\b1,\b1,0,0,0,-2)\nn
\ea

According to the RGE running results of Table~\ref{rt1}, there is
only one brane configuration, for the models under discussion,
that reproduces the weak mixing angle at low energies. This consists
of placing the $\u1_1$ brane on the top of the color branes, so that
$g_1=g_3$, which leads to $M_s\sim{\cal O}(10)$ TeV (7-17 TeV, depending on the
threshold corrections).

\section{Fermion masses}

Although the general question of quark and lepton masses goes beyond the scope of
this paper, we would like to make here some comments in the context of our
constructions. The Yukawa couplings relevant to fermion masses  are constrained by the
various U(1) symmetries and can present interesting patterns.

\bigskip
$\bullet$ {\bf Model A}. The relevant Yukawa couplings are
\ba
M_A=\lambda_u\,Q\, u^c\, H_u +\lambda_d\,Q\,d^c\,H_d^\dagger+
\lambda_e\,L\,e^c\,H_u^\dagger+\lambda_\nu\,L\,H_d\,\nu_R
\ea
Here, charged
leptons and up quarks (of the heaviest generation) obtain masses from the same
Higgs ($H_u$).

When all Yukawa couplings arise at the lowest (disk) order, it is easy to check
that in the simplest case (absence of discrete selection rules, etc), they
satisfy the following relations:
\ba
\lambda_u =\lambda_e =\sqrt{2}g_2\ ,\quad \lambda_d =\sqrt{2g_s}\ ,\quad
\lambda_\nu =\sqrt{2}g_b\, .
\label{A}\ea
The top and bottom quark masses are given by:
\ba
m_t=g_2 v \sin\beta\qquad ;\qquad m_b={\sqrt g_s}v\cos\beta\, ,
\label{beta}\ea
where $\tan\beta=v_u/v_d$, with $v_u$ and $v_d$ the vacuum expectation
values (VEVs) of the two higgses $H_u$ and $H_d$, respectively, and
$v=\sqrt{v_u^2+v_d^2}=246$ GeV.
Note that in the case where the color branes are identified
with D3 branes, one has $\sqrt{g_s}=g_3$, and in any case $g_s\ge g_3^2$.
Note also that since the string scale in this model is relatively low, $M_s\simlt
1$ TeV, there is no much evolution of the low energy couplings from the
electroweak to the string scale. Thus, using the known value of the bottom mass
$m_b\simeq 4$ GeV, one obtains for the top quark mass $m_t\simeq 162$ GeV
which is less than 5\% below its experimental value
$m_t^{\rm exp}=174.3\pm 5.1$ GeV. In addition, the Higgs VEV ratio turns
out to be large, $\tan\beta\simeq 100$. Note that such a large value is
not in principle problematic as in the supersymmetric case, but it can
lead to important higher order corrections.

On the other hand the $\tau$-mass is of the same order as the top
mass, which is unrealistic.
However, there is still the possibility that
the lepton Yukawa coupling $\lambda_e$ vanishes to
lowest order due to additional string discrete selection rules, and is generated by a higher dimensional operator of
the form $Le^c(H^\dagger_u H^\dagger H)$ providing the appropriate
suppression.\footnote{Models with similar properties have been considered in the past
in the perturbative heterotic string framework.}

\bigskip
$\bullet$ {\bf Model A'}. The Yukawa couplings here are
\ba
M_{A'}=\lambda_u\,Q\, u^c\, H_u + \lambda_d\,Q\,d^c\,H_d^\dagger+
\lambda_e\,L\,e^c\,H_d^\dagger+\lambda_\nu\,L\,H_d\,\nu_R
\ea
with the same relation for the tree-level couplings as in
(\ref{A}).
Using the parametrization in (\ref{beta}) we see that the
relation of $m_{t}$ to $m_{b}$ is the same as in model A and the
same remarks apply.
Since here the lepton and down quark acquire their
masses from the same Higgs, one obtains the phenomenologically interesting
relation: $m_b/m_\tau =\sqrt{g_s}/g_2=g_3/g_2$, when strong interactions are on
D3 branes. Thus, from Table~\ref{rt1}, $m_b/m_\tau\simeq 1.75$ at the (string)
unification scale, which is in the upper edge of the experimentally
allowed region at the $Z$-mass, $1.46\simlt m_b/m_\tau|_{\rm exp} \simlt
1.75$. This relation could replace the successful GUT prediction
$m_b=m_\tau$ of the conventional unification framework, in low scale string
models.
In conclusion model A' seems to be able to generate the required
hierarchy of masses for the third generation.

\bigskip
$\bullet$ {\bf Model B}.
The relevant trilinear Yukawa couplings are,
\ba
M_B=\lambda_u\,Q\, u^c\, H_u + \lambda_d\,Q\,d^c\,H_d^\dagger+
\lambda_e\,L\,e^c\,H_d^\dagger+\lambda_\nu\,L\,H_u\,\nu_R
\ea
The tree-level Yukawa couplings satisfy
\ba
\lambda_e=\lambda_u =\sqrt{2g_s}\ ,\quad \lambda_d =\sqrt{2}g_3\ ,\quad
\lambda_\nu =\sqrt{2}g_b\
\ea
and we have
\ba
m_t={\sqrt g_s}v \sin\beta\qquad ;\qquad m_b=g_3 v\cos\beta\, .
\label{rgebd}
\ea

The first relation implies again a heavy top, while the bottom to tau mass
ratio is now predicted, with a value
$m_b/m_\tau=g_3/\sqrt{g_s}\simlt 1$ which is apparently far from its
experimental value. However, in this case, the string scale is
relatively high and therefore one should take into account the
renormalization group evolution above the weak scale.
Solving the associated RGEs with the boundary conditions (\ref{rgebd})
and assuming $g_3=\sqrt{g_s}$, we obtain acceptable $m_b$ and $m_\tau$
masses for $M_s\sim 3 \times 10^3$ TeV and $\tan\beta\sim 80$.
Note that the successful prediction of $m_b$ and $m_\tau$ is related to
the condition $m_b=m_\tau$ at the (string) unification scale,
which in the case of non-supersymmetric Standard Model is obtained at
relatively low energies~\cite{Arason:1991ic}.
%======changed Jan 03
\begin{figure}[t]
\center \epsfxsize=12cm \epsfbox{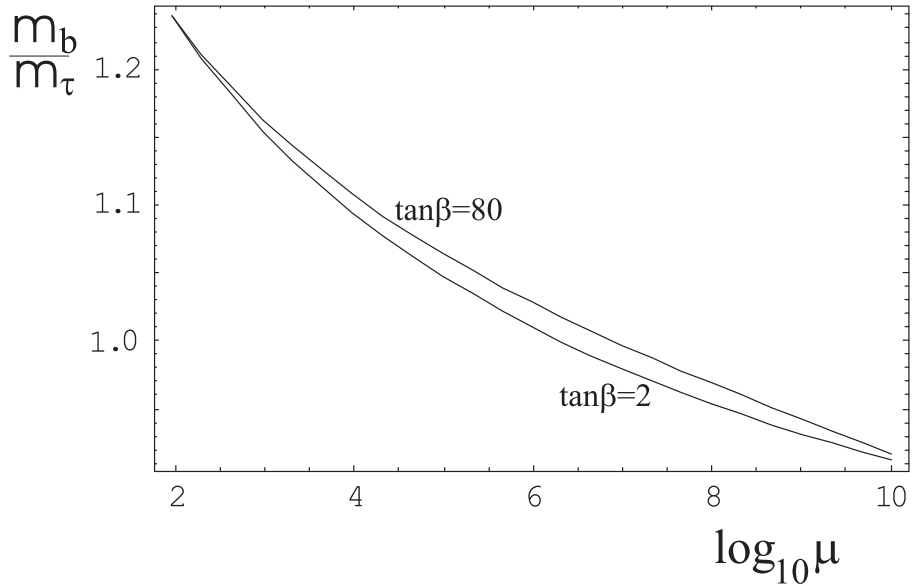}
\caption{\label{mbomtau}{\it Evolution of the ratio $m_b/m_\tau$
as a function of the energy $\mu$ for $\tan \beta=2$ and $\tan
\beta=80$. We have used as low energy parameters $m_b=4$ {\rm
GeV}, $m_{top}=174$ {\rm GeV} $a_3(M_z)=0.12$,
$\sin^2\theta_W=0.23113$.}}
\end{figure}
Indeed, in Figure~\ref{mbomtau}, we plot the mass ratio $m_b/m_\tau$ as a
function of the energy, within the non supersymmetric Standard Model with
two Higgs doublets.
Nevertheless, the resulting value of $M_s$ is still significantly
higher than the unification scale required from the analysis of gauge couplings
in section 3. Moreover, the top quark mass turns out to be rather high,
$m_t\sim 220$ GeV. It is an open question wether this discrepancy can be
attributed to threshold corrections that can be important in the case
of two dimensional bulk \cite{AB}.

\bigskip
$\bullet$ {\bf Model B'}.
The relevant Higgs couplings are given by
\ba
M_{B'}=\lambda_u\,Q\, u^c\, H_u +
\lambda_d\,Q\,d^c\,H_d^\dagger+\lambda_e\,L\,e^c\,H_u^\dagger+
\lambda_\nu\,L\,H_u\,\nu_R
\ea
while the tree-level Yukawa couplings by
\ba
\lambda_u =\sqrt{2g_s}\ ,\quad \lambda_d =\sqrt{2}g_3\ ,\quad
\lambda_\nu =\sqrt{2}g_b\ {\rm and}\ \lambda_e =\sqrt{2}g_2
\ea

Here, as in model A, the $\tau$ and top mass are of the same
order and thus in conflict with experiment. As in model A,
vanishing leading order coupling could be a way out.

In the above analysis we have also assumed that only the heaviest generation acquires
masses at the lowest order. The other two are considered to have vanishing
trilinear Yukawa couplings. This property does not follow from the gauge symmetries
we considered and should be attributed either to discrete string symmetries or
to additional gauge symmetries by enlarging the model.$^6$

\section{Neutrino physics}

One of the challenges of Standard Model extensions is the
justification of the smallness of neutrino masses. The favorite
scenario used to rely upon the introduction of
right-handed neutrinos (SM singlets) and their mixing with some
extra massive singlets. The suppression of the neutrino masses is
then obtained as a result of the structure of the full mass matrix
(``see-saw" mechanism). In order for this mechanism to work
effectively, the extra singlet mass should be about ten orders of
magnitude higher than the electroweak scale.

Among the promising features of D-brane models is a novel
scenario to account for  neutrino masses: right-handed neutrinos
are assumed to propagate in the bulk while left-handed neutrinos,
being a part of the lepton doublet, live on the brane. As a result,
the Dirac neutrino mass is naturally suppressed by the bulk
volume. Adjusting this volume, so that the string scale lies in the
TeV range, leads to tiny neutrino masses compatible with current
experimental data.

The extra dimensional neutrino mass suppression mechanism
described above can be destabilized by the presence of a large
Majorana neutrino mass term. As already mentioned in section 3,
the lepton-number violating  dimension five effective operator $L L H H$
leads, in the case of TeV string scale models, to a Majorana mass term
of the order of a few GeV. Even if we manage to eliminate this operator
in some particular model, higher order operators would also give
unacceptably large contributions, as we focus on models in which the
ratio between the Higgs vacuum expectation value (VEV) and the string scale
is just of order ${\cal O}(1/10)$. The best way to protect tiny neutrino
masses from such contributions is to impose lepton number conservation. As
we have seen in section 2, we can find models which successfully
accommodate all SM particles and preserve lepton number as an
effective global symmetry in perturbation theory. These are the models $A$,
$A'$ and $B$, $B'$ described in detail in section~3.

Apart from neutrino masses these theories contain also the
ingredients to explain neutrino oscillations. The right-handed
neutrino, being a bulk state, has a tower of Kaluza--Klein (KK)
excitations. Their mixing with the ordinary (left-handed) neutrino
leads to oscillation patterns that have to be compared with
present solar and atmospheric neutrino data. There exist extended
discussions in the literature \cite{bulknu} regarding the neutrino
mass and oscillation problems in the context of extra dimensional
theories. Among the common results of these works is that an
explanation of the solar neutrino anomaly is possible provided the
Small Mixing Angle (SMA) solution is acceptable. However, recent SNO results in
conjunction with SuperKamiokande data \cite{LND,SMI,ALT,ST} strongly
disfavor the SMA solution and thus render this higher
dimensional oscillation mechanism problematic, at least as far as solar
neutrino oscillations are concerned. A possible way out is to
introduce three bulk neutrinos and explain the oscillations in the
traditional way \cite{LANG}. The effect of the KK mixing can be
eliminated by  appropriately decreasing the size of the extra
dimensions and thus increasing the value of the string scale.
However, all these discussions are restricted to the case of
effectively one-dimensional bulk.
Besides these phenomenological difficulties, there is also a serious
theoretical problem, since one-dimensional propagation of massless
bulk states gives rise to linearly growing fluctuations which yield
in general large corrections to all couplings of the effective field
theory, destabilizing the hierarchy~\cite{AB}.

Two-dimensional scenarios
have not been considered in detail. We will see below how the above
problems can be resolved and discover that a two-dimensional bulk
has enough structure to describe both the solar and atmospheric
neutrino oscillations by introducing a single bulk neutrino pair.
On the other hand, recent experiments are also able to
differentiate between the contributions of active and sterile
neutrinos to the neutrino anomaly problems. From this point of
view, the KK excitations do not carry any Standard Model charges
and are thus considered as sterile. It is then important to
examine if all these constraints are compatible with our model.

As explained in the introduction, our setup incorporates a two-dimensional
bulk and we are going to assume hereby that neutrinos propagate in the full
bulk volume which is a two-dimensional space. Among the common features of the
models considered in section 3, one finds tree-level neutrino couplings and
mass-terms of the form:
\ba
\sum_{i=1}^3\lambda_i L_i H_i\nu_R\ \rightarrow\
\sum_{i=1}^3\lambda_i\,v_i\, \nu_{iL}\, \nu_R\, ,
\ea
where $i$
is a generation index and for each generation $i$, $H_i$ is one of the
available Higgs doublets $H_d$ or $H_u$, providing masses to
down quarks (models $A,A'$) or to up quarks (models $B,B'$), respectively,
with $v_i=\langle H_i\rangle$ the corresponding VEV.
The above couplings provide a mass to one linear combination
($\nu_{L}$) of the weak eigenstates ($\nu_{iL}$), while
the other two remain massless. Note, that it would be possible to
generate masses for all left-handed neutrinos by introducing
additional bulk neutrino pairs. In this case the number of
free parameters is increased and predictability is lost. Thus,
here, we will study the case of a single bulk neutrino pair.
Defining $N_L=(\nu_L,\nu_{0L},\nu_{0L}')$ the mass eigenstates, the
weak eigenstates can be written as
\ba
\nu_{iL}=\sum_j
U_{ij}\, N_{Lj}\label{two}\, ,
\ea
where $U$ is a $3\times3$
unitary matrix with $U_{j1}^*=(U^{-1})_{1j}=\frac{\lambda_j v_j}{m_D}$ and
$m_D^2=\sum_{i=1}^3\lambda_i^2 v_i^2$ is the mass-square of the
massive combination ($\nu_L$). Being of brane-bulk type, the couplings
$\lambda_i$ are naturally suppressed by the bulk volume $v_{89}$ (see
section 1) and lead to a tiny Dirac neutrino mass
\ba
m_D=\frac{\bar{v}}{\sqrt{{{v_{89}}}}}=\frac{2\sqrt{2}}{g_3 g_2} \frac{M_s}{M_P}
\bar{v}\ , \label{mddef}
\ea
where $\bar{v}=\sqrt{\sum_{i=1}^3h_i^2 v_i^2}$ with $h_i$, $i=1,2,3$, the
associated dimensionless Yukawa couplings and $v_i$ the corresponding
Higgs VEV $(v=\langle H_d\rangle,\langle H_u\rangle)$ depending on
the model. Using typical values for the gauge couplings
(see Table~\ref{rt1} of section 3), $v_i<v=246$ GeV and $h_i/4\pi=\ord{1}{}$,
we obtain $ m_D < 6 \times 10^{-3}\ {\rm eV}$
% previous \label{mdv}
for $M_s\simlt 10$ TeV. This provides an explanation for the
smallness of neutrino masses and
is actually the extra-dimensional version of the see-saw mechanism.
% assuming
%reasonable Yukawa coupling upper bound $\sqrt{\sum_{i=1}^3h_i^2
%v_i^2}\le 200GeV$ we obtain \ba m_D\le 7\times10^{-5}-7\times
%10^{-4} eV \label{mdrange} \ea {\tt I have to rewrite this to fit
%with the results at the end} for $M_s=1-10 TeV$.
% Of course, the above picture is simplified because we have neglected  the
%tower of Kaluza--Klein excitations.
%Only in the case where $\frac{1}{R_{8}},\frac{1}{R_9}>>m_D$ these excitations decouple,
%and  the above
%results, which take into account only the zero mode contributions
%$\nu_R=\nu_R^{(0)}$, are valid.

The above picture is simplified because we have neglected the contributions
of the tower of KK neutrino states. Taking them into account, and
assuming for simplicity that the two bulk radii
are equal $R_8=R_9=R$ and form an angle $\frac{\pi}{2}-\theta$,
where $-\pi/2<\theta<\pi/2$,
the mass terms become
\ba
L_{m}= m_D\,{\nu}_{L}\,\sum_{\vec{k}}\,\delta^{-\frac{m_{\vec{k}}^2}{M^2 }}
\,{\nu}^{(\vec{k})}_R +
\sum_{\vec{k}}m_{{\vec{k}}}\,{{{\nu}^c_R}^{(\vec{k})}}\,{\nu}^{(\vec{k})}_R
+c.c.
\label{mass}
\ea
where $\nu_R=\nu_R^{(0)}$ and the summation over $\vec{k}$ extends over all KK
%levels
momenta. By  $m_{\vec{k}}^2$ we denote
the mass-square of the KK excitation labelled by momenta $\vec{k}=(k_1,k_2)$
\ba
m_{\vec{k}}^2=\frac{1}{R^2 \cos^2\theta}\left(k_1^2+k_2^2-2 k_1 k_2
\sin\theta\right)\, .
\label{masfm}
\ea
We also use the notation $m_{k}^2$ for the mass-square of the $k$-th KK level.
$\delta$ is a model dependent constant bigger than one, associated to
the coupling of two Neumann--Dirichlet (ND) $Z_2$-twisted strings to an
untwisted (NN or DD) string~\cite{ttu}. In our models, there are four
$Z_2$-twisted coordinates, implying $\delta=16$.
$M$ plays the role of an ultraviolet (UV) cutoff, which is normally
the string scale $M_s$, but we prefer using the symbol $M$ because in certain
processes there exists an induced cutoff that can be a few orders of
magnitude below $M_s$. For instance, this is the case of solar neutrinos,
where the production energy is of order of a few MeV, and thus heavier KK
modes are effectively cut off.

The mass terms (\ref{mass}) lead to a mixture of the usual left-handed neutrino
with the infinite tower of its KK excitations. A detailed  analysis
of the eigenstate problem in our framework  is presented in Appendix A where
we derive the basic formulas for neutrino masses and transition probabilities.
Due to its complexity, the problem can be either treated numerically in the
general case or analytically using some approximation. The first approach
has the disadvantage of being rather tedious as it involves summations over a
very large number of KK modes, so we will adopt here an analytic perturbative
approach. Concerning the interpretation of neutrino anomalies, there are
also two possible treatments: The first is a direct fit of neutrino data to the
transition probability formulas obtained in our framework. The second is to
try to simulate the standard solutions to the solar and atmospheric neutrino
anomaly problems. We will use here the second method, as it is
sufficient for demonstrating the basic features of our model.

Following Appendix A, the mass spectrum of the full system, in the case of two
bulk dimensions, is:
\ba
\tilde{m}_k^2 = m_{k}^2+r_{{k}} m_D^2\,\delta^{-\frac{2 m_{k}^2}{M^2}}
\left(1-\Delta_k\right)+\dots
\ea
where $r_k$ is the multiplicity of the $k$-th KK level and
\ba
\Delta_k=\pi\,m_D^2\,(R^2\cos\theta)\,
\left\{\log\left(M^2\,R^2\cos\theta\log\delta^2\right)+ s_k\right\}\, ,
\label{deltanm}
\ea
with $s_k$ a volume independent constant. Our solution is based on
the assumptions that $m_DR\ll1$, as justified by (\ref{mddef}) and
$\Delta_n<1$ that simplifies the formulas involved. Under these assumptions,
and following the analysis in Appendix A, the survival probability for a neutrino of flavor $i$ is given by
\ba
 P_{\nu_i\to\nu_i} &\approx& 1 -4\,u_i^2(1-u_i^2)\,
\sin^2\left(\frac{m_0^2}{4}\frac{L}{E}\right)
-3.2\,u_i^2\Delta_0\,\sin^2\left(\frac{\omega}{R^2}\frac{L}{4
E}\right) \label{survam}\, , \ea
where $m_0^2=m_D^2(1-\Delta_0)$
and $u_i=|U_{i1}|$ satisfying the unitarity relation $\sum_i
u_i^2=1$; $L$ is the distance that the neutrino travels before
being detected and $E$ is the beam energy. According to the
discussion in Appendix A, the survival probability takes this form
only for specific values of the angle~$\theta$:
\ba
\sin\theta=\frac{p}{q}\ ,\quad p,q\in\mathbb{Z}\ ,\quad
|p|<q\qquad ;\qquad
 \omega=\frac{q(3+(-1)^q)}{2(q^2-p^2)}\, ,
\label{msinpq}
\ea
where $p,q$ are relatively prime integers and  $q=1$ for $p=0$.
%We will also assume for reasons that will become clear later,
%that we choose appropriately
%$p,q$ (e.g. $q=p+1$, $p\gg1$) so that $\omega$ is close to unity that is
%$\theta$ approaches $\tfrac{\pi}{2}$.
In our approximation, the survival probability (\ref{survam}) is a
superposition of two modes with frequencies:
$\frac{m_D^2 L}{4 E}$  and $\frac{\omega L}{4 R^2 E}$.
These two frequencies can be considered as independent parameters, as the
first depends on the Yukawa couplings and the Higgs VEVs, while
the second depends on the compactification radius or equivalently on the
string scale. The existence of these two frequencies provides us with the
opportunity to fit both solar and atmospheric oscillations using
(\ref{survam}). Furthermore, the amplitudes of the two modes depend on $u_i$
and $\Delta_0$ defined in (\ref{deltanm}). These parameters can be used
in order to fit the oscillation amplitudes.

In the standard neutrino (two flavor) scenario, one usually explains
the solar neutrino anomaly by $\nu_e\to\nu_\mu$ oscillations and
the atmospheric neutrino deficit by $\nu_\mu\to\nu_\tau$
oscillations. The formula for the transition probability is:
\ba
P_{\nu_i\to\nu_j}=\sin^2 2\theta_{ij}\sin^2\left(
\frac{\Delta m_{ij}^2}{4} \frac{L}{E}\right)\, ,\label{sof}
\ea
where $\Delta m_{ij}^2=m_i^2-m_j^2$ is the neutrino mass difference in the
case of two states mixing. Expressing $L$ in kms, $E$ in GeV and $\Delta
m^2$ in eV$^2$, the frequency $\Delta m_{ij}^2 \frac{L}{4E}$ takes the form
$1.27\times \Delta m_{ij}^2\, L/E$.

Recent analysis of  atmospheric neutrino data \cite{ALT} at $3\sigma$
C.L. gives
$1\times10^{-3}<\Delta m^2_{\rm atm} < 6\times 10^{-3} {\rm eV}^2$ and
$0.7<\sin^22\theta_{\rm atm}<1$. Regarding solar data, the situation has
dramatically changed after the latest SNO results: Only the LMA and LOW MSW
solutions are acceptable at the $3\sigma$ C.L. with $2.3 \times
10^{-5}<\Delta m^2_{\rm LMA}< 3.7\times 10^{-4} {\rm eV}^2$,
${0.6<\sin^2 2\theta}_{\rm LMA}<1$ and
$3.5 \times 10^{-8}<\Delta m^2_{\rm LOW}< 1.2\times 10^{-7} {\rm eV}^2$,
$0.8<{\sin^2 2\theta}_{\rm LOW}<1$. Moreover, the LMA gives a much better fit.
The region of the SMA solution
(with best fit values $\Delta m^2_{\rm SMA}\sim 5\times 10^{-6} {\rm eV}^2$,
${\sin^2 2\theta}_{\rm SMA}\sim 2\times 10^{-2}$) is acceptable only at the
$5.5\sigma$ level and is thus practically excluded~\cite{SMI}.

The atmospheric neutrino oscillation frequency is higher than all
solar ones, $\Delta m^2_{\rm atm} > \Delta m^2_{\rm
sol}$, and thus we have to use the lowest frequency in
(\ref{survam}) (i.e $m_0^2$) to simulate solar neutrino
oscillations. Formula $(\ref{survam})$ contains four independent
parameters, namely $m_D, R, M_s$ and $u_e$ (assuming $u_\tau=0$
and thus $u_e^2+u_\mu^2=1$). Fitting both solar and atmospheric
oscillations requires to leading order in $\Delta_0$: \ba
\frac{\omega}{R^2}&=&{\Delta m}_{\rm atm}^2 \label{efa}\\
m_D^2&=&{\Delta m}_{\rm sol}^2
\label{efb}\\
4\,u_e^2(1-u_e^2)&=&\sin^22\theta_{\rm sol}\label{efc}\\
3.2\,u_{\mu}^2\,\Delta_0&=&\sin^22\theta_{\rm atm}
\label{efd}
\ea
Neglecting the constant term $s_0$ in the expression (\ref{deltanm}) of
$\Delta_0$, in the limit $M R\gg1$, and assuming $\delta=16$, $\Delta_0$ can
be written in terms of $M_s$ and $m_D$ as
\ba
\Delta_0 \approx \frac{1}{2 \pi a^2}\frac{m_D^2 M_P^2}{M_s^4}
\log\left(\frac{M_P }{ \pi a M_s}\sqrt{\frac{\log\delta}{2}}\right)\quad ;\quad
a=\frac{2 \sqrt{2}}{g_3 g_2}\, ,
\label{adelta}
\ea
where we have assumed that the cutoff is equal
to the string scale ($M=M_s$). This choice of the
cutoff is suitable for the atmospheric neutrino data, where the oscillation
amplitude is proportional to $\Delta_0$. In any case, the exact value of the
cutoff plays a minor role in our calculation, due to the fact that it
appears always logarithmically.
Furthermore, the expectation value $\bar{v}$ is related to the rest of the
parameters through equation (\ref{mddef}), while the angle $\theta$ enters
in the Plank mass definition (\ref{mp})
\ba
R^2 \cos\theta=\frac{1}{4 \pi^2 a^2} \frac{M_P^2}{M_s^4}\, .
\label{vbulk}
\ea

In terms of the integers $p,q$ that have been introduced in
(\ref{msinpq}), we can recastrewritw the last equation as \ba
W_{p,q}\equiv\frac{3+(-1)^q}{2 \sqrt{q^2-p^2}}= \frac{\Delta m_{\rm
atm}^2}{4 \pi^2 a^2} \frac{M_P^2}{M_s^4}\, , \label{wpq} \ea or
equivalently \ba \cos\theta=\frac{3+(-1)^q}{2 W_{p,q}\, q}\, . \ea
Thus, the four conditions (\ref{efa})-(\ref{efd}) together with
(\ref{adelta}) and (\ref{mddef}), (\ref{vbulk}) fix all four
parameters of the model. Therefore, fitting the atmospheric
neutrino frequency (\ref{efa}), one determines the
compactification radius \ba 1 \times 10^{-3} {\rm
eV}^2<\frac{\omega}{R^2}<6\times 10^{-3} {\rm eV}^{2}\, ,
\label{atmrange} \ea or $3\ \mu{\rm m}<R<6\ \mu{\rm m}$ for
$\omega\sim 1$. Choosing for the solar neutrino deficit the
preferred LMA solution, we get from the second condition
(\ref{efb}) the neutrino mass range: \ba 4.8 \times 10^{-3} {\rm
eV} < m_D < 7.7\times 10^{-2} {\rm eV}\, . \label{mdlma} \ea The
third condition (\ref{efc}) fixes the mixing coefficient $u_e^2$
and has two possible solutions, namely, $0.18<u_e^2 <0.5$ or
$0.5<u_e^2<0.82$. Choosing $u_e^2\simeq 0.18$ and $u_{\mu}^2\simeq
0.82$ ($u_\tau=0$), equation (\ref{efd}) leads to $\Delta_0\sim
0.27$ (in the case we choose the lowest allowed value of
$\sin^22\theta_{\rm atm}$), which lies at the edge of the validity
of our perturbative approach. Any other choice of $u_i$ compatible
with the constraints leads to bigger values for $\Delta_0$. This
justifies also the choice $u_\tau=0$ in order to minimize
$\Delta_0$ in (\ref{efd}).\footnote{Normally, one should repeat
the eigenstate analysis of Appendix A numerically in the
non-perturbative region, but from a preliminary analysis we do not
expect significant change of our results.} From (\ref{efd}) we get
the string scale: \ba 8\, {\rm TeV} \simlt M_s \simlt 13\, {\rm
TeV}\, , \ea while compatibility with (\ref{mddef}) requires
$\ord{1}{}$ values for the Yukawa couplings. It is interesting
that this range for the string scale coincides with the values we
found from the analysis of gauge couplings in section 3, for the
models $B$ and $B'$. Coming to the angle, we get from (\ref{wpq})
$0.02\lsim W_{p,q}\lsim 0.2$ for the allowed range of $\Delta
m_{\rm atm}^2$ and we can easily verify that there exist integers
$p,q$ that satisfy (\ref{wpq}).

Let us now consider the LOW solution to the solar neutrino deficit.
Following similar steps, the four constraints (\ref{efa})-({\ref{efd}) in
this case give
\ba
1.9 \times 10^{-4} {\rm eV}< m_D <3.5 \times 10^{-3} {\rm eV}\, ,
\ea
with $u_e^2\approx 0.28$ and thus $u_\mu^2\approx 0.72$ and $\Delta_0\approx
0.30\,$. The string scale turns out to be  slightly lower in this case:
\ba
1.8\, {\rm TeV} \simlt M_s \simlt 2.2\, {\rm TeV}\, ,
\ea
while for the angle $\theta$ we get $20 \lsim W_{p,q} \lsim 200$.
Note that the range of string scale is now compatible with the values found
from the analysis of gauge couplings in models $A$ and $A'$. In this
solution, the left-handed neutrino Yukawa couplings do not have to be of
$\ord{1}{}$.

Moreover, the practically excluded  SMA solution can also be obtained in this
framework. The associated parameters in this case are:
$u_e^2\sim 5\times 10^{-3}$, $u_\mu^2=1-u_e^2\approx 1$, $\Delta_0\sim 0.2$,
$m_D\sim2\times 10^{-3}{eV}$, $M_s\sim 6$ TeV, $0.2\simlt W_{p,q}\simlt 1.2$.
Note that the case $\theta=0$ corresponds to $p=0,q=1$ and thus  $W_{p,q}=1$.
As seen from our results, only the SMA solution includes this value in the
allowed $W$-range and this is the reason why only this solution could be
reproduced in the case of an orthogonal torus. The LMA and LOW solutions
require a bulk forming a non-orthogonal lattice, corresponding to
non-trivial values of $\theta$. It is also worth noticing that such
non-trivial values of $\theta$ induce CP violation in the neutrino sector,
which is interesting to be further explored.

The mixing of the neutrino zero mode with its KK excitations can lead to a
decay of the left-handed neutrino to these KK modes, considered as sterile
from the SM point of view. In our framework, and to leading order in the
$\Delta_0$ expansion, the average conversion rate of a
neutrino of flavor $i$ to sterile is given by (\ref{pster}):
\ba
\bar{P}_{\nu_i\to s}\sim
2 u_i^2 \Delta_0\ .
\ea
Constraints (\ref{efb}) and (\ref{efd})
fix both the above probabilities. Assuming the LMA solution to the solar
neutrino deficit, we get
\ba
\bar{P}_{\nu_\mu\to s}\sim 0.44\
\ea
for atmospheric
and
\ba
\bar{P}_{\nu_e\to s}\sim 0.05\
%; \ \bar{P}_{\nu_e\to \nu_\mu}\sim
%0.20
\ea
for solar neutrinos, where in the second case we have assumed
a cutoff $M\sim 50$ MeV. For the LOW solution, the transition probabilities
are similar: $\bar{P}_{\nu_\mu\to s}\sim 0.32$, $\bar{P}_{\nu_e\to s}\sim
0.08$.
Note that the decay rate to sterile neutrinos is significant
in the case of atmospheric neutrinos and is negligible in the case of
solar neutrinos. This is related to the structure of our model for neutrino
oscillations. The atmospheric neutrino deficit is simulated using
the lightest KK neutrino excitation (which is interpreted from the SM point of
view as a sterile neutrino), while the solar data are explained using
the (active form the SM point of view) zero mode.

%===========================================================
Constraints for the conversion of active to sterile neutrinos have
been recently examined in reference \cite{LANG}. Following their
analysis in the case of the LMA solution, the constraint to the
average decay rates for solar neutrinos is $\bar{P}_{\nu_e\to
\nu_s}<0.40$ at 90\% c.l. which is obviously satisfied by our
model. For atmospheric neutrinos the relative constraint takes the
form
\ba
\Delta P=\bar{P}_{\nu_\mu\to \nu_s}-\bar{P}_{\nu_e\to
\nu_s}<0.17\ .\label{ccons}
\ea
Evaluating this constraint in our framework, one finds
$\Delta P= 0.44-0.10=0.34$ (where $\bar{P}_{\nu_e\to \nu_s}=0.10$ in the
case of atmospheric due to the higher cutoff in $\Delta_0$ of eq (\ref{deltanm}))
which is  by a factor of two higher than the experimental bound.
%Futhermore, strictly speaking the constraint (\ref{ccons}) applies to
%the sub-GeV region. Repeating our calculation with cutoff $M=1GeV$ we get
%$\Delta P\sim 0.32-0.07=0.25$.
However, one should take into consideration that our perturbative
analysis, focusing on explicitly revealing oscillations, does not allow
to access the region $\Delta_0\sim1$ where in principle the above rates
could change. As mentioned earlier this region could be studied only numerically.
This requires summation over a huge number of KK modes and at present
it appears insoluble even numerically.

In any case the exact nature of atmospheric neutrino oscillations is expected to be
further examined in the K2K \cite{K2K} experiment. In case the predictive
scenario of a single bulk neutrino presented here fails to satisfy the sterile
production constraints,
one should proceed in the introduction of additional
bulk neutrinos and explain oscillations in the traditional way, that
is by zero mode mass difference and not by mixing with the KKs.
Their presence can still lead to sterile production which
can be reduced  by appropriately raising the string scale and thus
decoupling the KKs \cite{LANG}.

\section{Summary and conclusions}

In conclusion, we performed a systematic study of the Standard Model
embedding in type I string theory at the TeV scale. We found that the
minimum configuration with interesting phenomenological features requires
three sets of D-branes, so that all SM particles are obtained as open
strings stretched among these brane stacks. Two of them describe
respectively the strong and weak interactions, while the third one
contains a single abelian brane that extends in a two-dimensional bulk of
submillimeter size.

The model predicts the correct value of the weak angle for a string scale of
a few TeV. It also contains baryon and lepton number as perturbative global
symmetries, ensuring proton stability and absence of large (Majorana)
neutrino masses. On the other hand, it uses two Higgs doublets that can
provide masses to all quarks and leptons. Concentrating on the heaviest
generation, we computed all trilinear Yukawa couplings and studied the
resulting mass relations. We found a naturally heavy top and the mass ratio
of bottom quark to tau lepton close to its experimental value.

Finally, we have studied neutrino masses and oscillations by introducing a
single right-handed neutrino state in the bulk. We found that both solar and
atmospheric neutrino data can be explained if the bulk is a non orthogonal
torus forming a non-trivial angle. Solar oscillations are then explained
using the zero-mode, which obtains a tiny mass from the electroweak Higgs,
while atmospheric oscillations use its first KK excitation. However,
in the cases of atmospheric data, it seems to be an excess in
sterile production with respect to current atmospheric data analyses.

Overall, the model looks very promising and deserves further investigation.
Particular directions that have not been discussed are the masses and mixing
angles of the two lightest generations, possible important threshold
corrections related to the two-dimensional bulk, supersymmetry breaking
effects in models with brane supersymmetry breaking, as well as explicit type
I string realizations.

\section*{Acknowledgments}

This work was partly supported by the European Union under the RTN
contracts HPRN-CT-2000-00148, HPRN-CT-2000-00122,
HPRN-CT-2000-00131, HPRN-CT-2000-00152, HPMF-CT-2002-01898 and the
INTAS contract N 99 1 590. E.K., J.R. and T.N.T. would like to
thank the Theory Division of CERN, J.R. the CPHT of Ecole
Polytechnique and T.N.T. the L.P.T. of the Ecole Normale
Sup\'erieure for their hospitality. I.A. would like to thank Guido
Altarelli for enlightening discussions.

%\newpage

\appendixA{Appendix A: Neutrino masses and oscillations}

We consider the neutrino mass eigenvalue problem that arises when
the usual left-handed neutrino localized on the ``brane" mixes with
one pair of right-handed neutrinos propagating in a
two-dimensional bulk. The solution of this problem in the case of
one-dimensional bulk has already been studied in the literature
\cite{bulknu}. A new feature of the two-dimensional bulk is
that the associated KK sums are divergent and a mass scale $M$
playing the role of the UV cutoff, normally identified with $M_s$,
appears in the mass and eigenstate expressions. Moreover, we
consider neutrino oscillations and  derive formulas for the
transition rates of both active and sterile neutrinos.

We will assume  for simplicity that the two bulk radii are equal,
$R_8=R_9=R$, but we will allow for the possibility that the angle
$\theta$ between the two compactified directions is arbitrary
$-\pi/2<\theta<\pi/2$. The masses of the KK excitations, labelled by
momenta $\vec{n}=(n_1,n_2)$, are:
\ba
m_{\vec{n}}^2=\frac{1}{R^2 \cos^2\theta}\left(n_1^2+n_2^2-2 n_1 n_2
\sin\theta\right)\, .
\label{masf}
\ea
The KK modes can be ordered according to their mass and
labelled by a unique level number $k$. Massive levels have in
general degeneracy four, apart from particular points that
have higher degeneracy for special values of $\theta$. In any
case, only the direct sum of the states of each degenerate level couples
to the left-handed neutrino. Hence, we can diagonalize in the
degenerate subspace and choose one of the eigenstates, which
corresponds to the sum of the degenerate KK modes. In this basis,
the relevant neutrino mass terms take the form
\ba
L_m=m_{D}\,{\nu}_{L}\,\sum_{k}\sqrt{r_k}\delta^{-\frac{{m_k}^2}{M^2}}\,
\tilde{\nu}^{(k)}_R + \sum_{k}m_{{k}}\,{{\tilde{\nu}}}^{c({k})}_R\,
\tilde{\nu}^{({k})}_R +c.c.+{\rm decoupled}\, ,
\label{massn}\nn
\ea
where
${\tilde{\nu}}_{R}^{({k})}=\frac{1}{\sqrt{r_{{k}}}} {\sum_{{\ell}}}
{\nu}_R^{({\ell})}$ and ${\tilde{\nu}}_{R}^{c
({k})}=\frac{1}{\sqrt{r_{{k}}}}\sum_{\ell}
{{\nu}^{c}}_R^{({\ell})}$, $\ell=1,\dots,r_k$, with
$r_{k}$ the multiplicity of the KK level with mass $m_k$. The mass
terms can be written in matrix form ($N^T_{L} m N_R$ +c.c.) with
$N_{L}=\left({\nu}_{L},{\tilde{\nu}}_R^{c (1)},\dots\right)$,
$N_{R}=\left(\tilde{\nu}_{R}^{(0)},{\tilde{\nu}}_R^{c
(1)},\dots\right)$ and $m$ an infinite-dimensional matrix.
%\ba
%m=\frac{1}{R}\left(\begin{array}{cccccc} x &\sqrt{p_{{n}_1}}
%\delta^{-\frac{{n}_1^2}{M^2 R^2}} x &\sqrt{p_{{n}_2}}
% \delta^{-\frac{{n}_2^2}{M^2 R^2}}x
%&\dots&\sqrt{p_{{n}_k}} \delta^{-\frac{{n}_k^2}{M^2 R^2}} x\\
%0            &{{n}_1}        &0         &\dots&0\\
%0            & 0                 &n_2&0&\dots\\
%\vdots       &\vdots             &\vdots     &\dots&0\\
%0            & 0                 &\dots&0&{{{n_k}}}\\
%\end{array}\right)
%\ea

In order to determine the left-handed neutrino mass eigenstates, we consider
\ba
&&m m^\dagger=\nn\\
&&\left(\begin{array}{cccccc} m_D^2 A& m_1 m_D\sqrt{r_1}\delta^{-\frac{m_1^2}{M^2}}&
m_2 m_D\sqrt{r_2}\delta^{-\frac{m_2^2}{M^2}}&
\dots&m_k m_D \sqrt{r_k} \delta^{-\frac{m_k^2}{M^2}}\\
m_1 m_D \sqrt{r_1} \delta^{-\frac{m_1^2}{M^2}} &m_1^2   &0    &\dots&0\\
m_2 m_D \sqrt{r_2} \delta^{-\frac{m_2^2}{M^2}} & 0 &m_2^2&\dots&0\\
\vdots       &\vdots             &\vdots&\vdots&\vdots\\
  \sqrt{r_{k-1}} m_{k-1} m_D
 \delta^{-\frac{m_{k-1}^2}{M^2}}          &0               &0&\dots&0\\
  \sqrt{r_{k}} m_k m_D
 \delta^{-\frac{m_k^2}{M^2}}          & 0                 &0&\dots&{m_k^2}\\
\end{array}\right)\nn\\
\ea
where $A=\sum_{\ell}r_{\ell} \delta^{-\frac{2m_{\ell}^2}{M^2}}$. In the
sequel we will assume that all masses are measured in string
units, which we restore only at the end of our calculations.

The exact eigenvalue equation for the mass of the $n$-th KK level
$\tilde{m}_{n}$ can be written in the form
\ba
m_D^2\,\sum_{{\ell}}
\frac{r_{\ell}\,\delta^{-2 m_{{\ell}}^2}}{\tilde{m}_{{n}}^2-{m}_{{\ell}}^2}=
%m_D^2\,\sum_{\vec{\ell}}
%\frac{\delta^{-2 m_{\vec{\ell}}^2}}{\tilde{m}_{\vec{k}}^2-{m}_{\vec{\ell}}^2}=
1\label{eigen}
\ea
and the associated eigenstates
\ba
\nu_L^{n}=\frac{1}{N_n}\,\left(1,c_1^n,c_2^n,\dots\right)
\ea
with
\ba
{c^{n}_\ell}=\frac{m_D m_\ell}{\tilde{m}_n^2-{m_{\ell}^2}}\,
 \delta^{-m_\ell^2}
\ea
and
\ba
N^2_{n}=m_D^2 \tilde{m}_{n}^2\,\sum_{\ell}
\frac{r_\ell\,\delta^{-2 m_\ell^2}}{(\tilde{m}_n^2-m_{\ell}^2)^2}\, .
\label{ndef}
\ea
The above results can be used to express $\nu_L$ in the basis
of the mass eigenstates
\ba
\nu_{L}=\sum_{n} \frac{1}{N_{n}}\,
\nu_L^{n}
\ea
and calculate its  time evolution
\ba
\nu_{L}(t)=\sum_{n} \frac{1}{N_{n}}\,\exp\left(\frac{i \tilde{m}_n^2 L}{2
E}\right)\,\nu_L^{n}\, ,
\ea
where $E$ is the neutrino beam energy and L is the
distance from the source. Therefore, using (\ref{two}) we can
derive the time evolution of the weak eigenstates
\ba
\nu_{iL}(t)&=&U_{i1} \nu_L(t)+U_{i2} \nu_{0L} +U_{i3} \nu_{0L}'\, .
\ea

The transition rate $P_{\nu_i\to\nu_j}$, that gives the
probability for a neutrino of a specific flavor $i$ produced in
the source to be detected as flavor $j$ in the detector, is
\ba
P_{\nu_i\to\nu_j}=\left|\langle\nu_{iL}(0)|\nu_{jL}(t)\rangle\right|^2=
\left\{\begin{array}{l}
\left|1-u_i^2+u_i^2 T\right|^2\ \ ,i=j\\
u_i^2 u_j^2\left|1-T\right|^2\ \ ,i\ne j
\end{array}
\right.\label{survij}
\ea
where $u_i=|U_{i1}|$ and
\ba
T\equiv\sum_n\frac{1}{N_{n}^2}\,\exp\left(\frac{i \tilde{m}_n^2 L}{2
E}\right)\, .
\label{sdef}
\ea
The formulas for the transition probabilities to active neutrinos are
\ba
P_{\nu_i\to\nu_j}
%={\left|1+u_i^2 (S-1)\right|}^2
&=&
\left\{\begin{array}{l}
(1-u_i^2)^2+u_i^2(1-u_i^2) \left({T+T^\ast}\right)+
 u_i^4 \left|T\right|^2\ ,\ i=j\\
 u_i^2 u_j^2\left(1-(T+T^\ast)+|T|^2\right)\ ,\ i\ne j\, .
 \end{array}
 \right.\label{np}
\ea
Therefore, the transition rate for a neutrino of flavor $i$ to decay into a
sterile neutrino is:
\ba
P_{\nu_i\to s}=1-\sum_{j=1}^3 P_{\nu_i\to \nu_j}=u_i^2(1-|T|^2)\, .
\label{nts}
\ea
Using (\ref{sdef}) we obtain
\ba
\frac{T+T^*}{2}&=&1-\frac{2}{N_0^2} \sin^2\frac{\tilde{m}_0^2 L}{4E}-2 F_2(\tilde{m}_n^2)\label{tptb}\\
{\left|T\right|}^2&=&1-4\,F_2(\tilde{m}_n^2)+4
F_2^2(\tilde{m}_n^2-\tilde{m}_0^2)+F_1^2(\tilde{m}_n^2)\label{tsq}
%\frac{4}{N_0^2}\sum_{n\ne0}\frac{1}{N_n^2}\sin^2\frac{(m_n^2-m_0^2)L}{4E}
%
%-2\sum_{n,k\ne0}\frac{1}{N_n^2 N_k^2}\sin^2\frac{(m_n^2-m_k^2)L}{2E}\label{tsq}
\ea
where
\ba
F_p(\tilde{m}_n^2)=\sum_{n\ne0}\frac{1}{N_n^2}{\sin^p \frac{\tilde{m}_n^2 L}{4 E}}\, .
\ea
We can now calculate the average  probabilities
for a neutrino of flavor $i$ to survive
\ba
\bar{P}_{\nu_i\to\nu_i}=(1-u_i^2)^2+u_i^2 \zeta^2\,,
\label{cnuinui}
\ea
or to be converted to flavor $j$
\ba
\bar{P}_{\nu_i\to\nu_j}=u_i^2 u_j^2(1+\zeta^2)\,,\ i\ne j\,,
\label{cnuinuj}
\ea
or to decay into sterile
\ba
\bar{P}_{\nu_i\to s}=u_i^2\left(1-\zeta^2\right)\, ,\label{pster}
\ea
%\ba
%\bar{P}_{\nu_i\to s}=2 m_D^2 u_i^2
%\sum_{\vec{n}\in\mathbb{Z}^2}\frac{\delta^{-{2 m_{\vec{n}}^2}}}{m_{\vec{n}}^2}=
%2 u_i^2 \Delta_0\, ,\label{pster}
%\ea
where
\ba
\zeta^2\equiv\sum_n\frac{1}{N_n^4}\label{dzeta}
\ea
and we have averaged over all frequency modes.

For $m_D=0$, the eigenvalues of the matrix $m m^\dagger$ form the usual KK
tower with masses $m_{\vec{n}}$ given by eq.~(\ref{masf}). For
$m_D R\ll1$, it is natural to assume that the KK levels are slightly
shifted:
\ba
\tilde{m}_{{n}}^2=m_{{n}}^2+\delta m_{{n}}^2\, .
\label{man}
\ea
Inserting (\ref{man}) into (\ref{eigen})
%and replacing summation over mass-levels ($\ell$)
%with summation over $\vec{\ell}$
and expanding for $\delta m_n^2 \ll m_n^2-m_\ell^2\ \forall\ n\ne\ell$, we
obtain to lowest order:
\ba
\delta m_n^2= \frac{r_n\,m_D^2 \delta^{-2 m_n^2}}{1+\Delta_n}\, ,
\ea
where
\ba
\Delta_n=m_D^2\,\sum_{\ell\ne n}\frac{r_\ell\,\delta^{-2
m_{\ell}^2}}{m_{\ell}^2-m_{n}^2}=m_D^2\sum_{\substack{\vec{\ell}\in\mathbb{Z}^2\\
m_{\vec{\ell}}\ne m_{{n}}}}
\frac{\delta^{-2 m_{\vec{\ell}}^2}}{m_{\vec{\ell}}^2-m_{{n}}^2}\, .
\label{deltan}
\ea
Due to the presence of the factor $\delta^{-2 m_n^2/M^2}$ (after restoring
the $M$ units), we have $\delta m_n^2\sim0$ for $m_n>M$, implying that KK
levels above the cutoff $M$ are not shifted. We notice also that in
(\ref{masf}) we have
$\frac{1}{R^2\cos\theta}\sim\frac{M_s^4}{M_P^2}\ll M_s^2$.
In this limit, we can calculate the leading terms in $\Delta_n$:
\ba
\Delta_n=\pi\,m_D^2\,(R^2\cos\theta)\,\left\{\log\left(M^2\,
R^2\cos\theta\log\delta^2\right)+ s_n\right\}\, ,
\label{deltann}
\ea
where $s_n$ is a constant term independent of the cutoff. Specifically,
$s_0=C$ and $s_{n\ne0}=-\log\left|n_1^2+n_2^2-2 n_1
n_2\,\sin\theta\right|+C$, with
$C$ a constant of order one. Similarly, for the normalization coefficients
we get to the lowest order in $\delta m_n^2$
\ba
\frac{1}{N_0^2}&=& \frac{\delta m_0^2}{m_D^2}+\dots\\
\frac{1}{N_n^2}&=&\frac{\left(\delta m_n^2\right)^2}{m_D^2 m_n^2 r_n \delta^{-2
m_n^2}}+\dots
\ea

The infinite  KK sums $F_2$, $F_1$ in (\ref{tptb}), (\ref{tsq})
can in principle be calculated numerically using (\ref{eigen})
and (\ref{ndef}). However, this requires summation over a huge
number of KK modes. A convenient approximation is to assume
$\Delta_n<<1$ for $n>0$. In this region
\ba F_2\sim
m_D^2\,\sum_{n\ne0} \frac{r_n\,\delta^{-2
m_n^2}}{m_n^2}\,\sin^2\left(\frac{m_n^2}{4}\frac{L}{E}\right)
=m_D^2 \sum_{\substack{\vec{n}\in\mathbb{Z}^2\\\vec{n}\ne(0,0)}}
\frac{\delta^{-2 m_{\vec{n}}^2}}{m_{\vec{n}}^2}\,\sin^2
\left(\frac{m_{\vec{n}}^2}{4}\frac{L}{E}\right)\, , \label{fsum}
\ea
and thus, $F_2$ can be expressed in terms of theta-functions
using the formula
\ba \sum_{\vec{n}\in\mathbb{Z}^2} \delta^{- 2
m_{\vec{n}}^2} \frac{\sin^2\left(\beta
m_{\vec{n}}^2\right)}{m_{\vec{n}}^2}&=&\mbox{Im}
\int_0^\beta dx\,\Bigl\{ \vartheta_3\left(-\frac{4 (x \log \delta
+i) }{\pi (1-\sin\theta) R^2}\right)
 \vartheta_3\left(-\frac{4 (x \log \delta + i)}{\pi (1+\sin\theta) R^2}\right)\nn\\
&~&+ \vartheta_2\left(-\frac{4 (x \log \delta + i) }{\pi (1-\sin\theta) R^2}\right)
 \vartheta_2\left(-\frac{4 (x \log \delta + i) }{\pi (1+\sin\theta) R^2}\right)\Bigr\}\, ,
\ea
where
$\vartheta_3(\tau)=\sum_{n\in\mathbb{Z}} e^{i\pi\tau n^2}$ and
$\vartheta_2(\tau)=\sum_{n\in\mathbb{Z}} e^{i\pi\tau
\left(n+\frac{1}{2}\right)^2}$. In the case of an orthogonal torus
($\theta=0$), $F$ is periodic under $\frac{L}{4 E}\to\frac{L}{4
E}+\pi\,R^2$. However, this property is lost for arbitrary values
of  $\theta$. The periodicity, of the survival probability
(\ref{survij}), is necessary for interpreting  the neutrino
anomaly through neutrino oscillations and is in general restored
for rational values of the angle $\theta$ \ba
\sin\theta=\frac{p}{q}\ ,\quad |p|<q\label{sinpq}\, , \ea where
$p,q$ are relatively prime integers, and $q=1$ for $p=0$.
Restricting $\theta$ to this subspace, $F$ becomes periodic under
$\frac{L}{4 E}\to\frac{L}{4 E} +\tau_{p,q}\pi R^2$, where \ba
\tau_{p,q}=\left\{\begin{array}{l}
\dfrac{q^2-p^2}{q}\quad \mbox{for}\ q=\mbox{odd}\\
\dfrac{q^2-p^2}{2q}\quad \mbox{for}\ q=\mbox{even.}
\end{array}\right.
\ea

Since $F_2$ is a periodic function, the next question is to compute the
corresponding amplitude. To get an estimation of the amplitude we
can evaluate the sum (\ref{fsum}) at the half period. We get
\ba
F_2^{1/2}=m_D^2 \sum_{\vec{n}\in\tilde{\mathbb{Z}}^2}
\frac{\delta^{- 2 m_{\vec{n}}^2}}
{m_{\vec{n}}^2}= \kappa \Delta_0\, ,
\ea
where $\tilde{\mathbb{Z}}^2$ is one of the sets of (even, odd) and/or
(odd, odd) integers of $\mathbb{Z}^2$, depending on the choice of $q$.
More particularly, for $q=4\,\ell$ with $\ell$ integer,
$\tilde{\mathbb{Z}}^2$ is the set of (odd, odd) pairs, for $q=2\,\ell+1$,
$\tilde{\mathbb{Z}}^2$ is the set of (odd, even) and (even, odd) pairs and
for $q=4\,\ell+2$, $\tilde{\mathbb{Z}}^2$ is the union of the two previous
sets. The constant $\kappa$ takes approximately the values
$\kappa\in\left(1/4,1/2,3/4\right)$ for each of the three cases,
respectively. Moreover, an upper bound to the amplitude of $F$ can be
derived by replacing the ``$\sin^2$" terms with unity
\ba
F_2^{\rm max}=m_D^2 \sum_{\vec{n}\in{\mathbb{Z}}^2}\frac{\delta^{- 2 m_{\vec{n}}^2}}
{m_{\vec{n}}^2}= \Delta_0\, .
\ea
Hence, the oscillation amplitude $\rho$
lies in the range $F_2^{1/2}<\rho<F_2^{max}$, that is $\kappa
\Delta_0<\rho<\Delta_0$ with $\kappa\in\left(1/4,1/2,3/4\right)$.
Furthermore, we can proceed to a numerical evaluation of $F$ for given
$\sin\theta$ and $M R$. An explicit example is presented in Figure~\ref{figf},
where we have calculated
$F$ as a function of $\frac{L}{4 E R^2}$ in the case
${M^2 R^2}=10^{4}$, $\delta=16$, $\sin\theta=\frac{119}{120}$.
In the same figure, we have also plotted the function
$\sin^2\left(\frac{240}{239} \frac{L}{4 E R^2}\right)$ (gray line) with
amplitude arising from the fit to the numerical sum data. The numerical
evaluation of the sum shows that the amplitude $\rho$ can in general be
approximated by $\rho\approx 0.8 \,\Delta_0$.
Taking into account these results, we will assume in section 6 that
the KK sum $F$ can be simulated by  the dominant frequency mode
\ba
F_2\approx
0.8\,\Delta_0\,
\sin^2\left(\frac{\omega_{pq}}{R^2}\frac{L}{4 E}\right)\quad ;\quad
\omega_{pq}=\frac{q(3+(-1)^q)}{2(q^2-p^2)}\, ,
\label{aproxr}
\ea
where $\Delta_0$ is given in (\ref{deltann})
and $\theta$ is given in (\ref{sinpq}).
\begin{figure}
\center
\epsfxsize=10cm \epsfbox{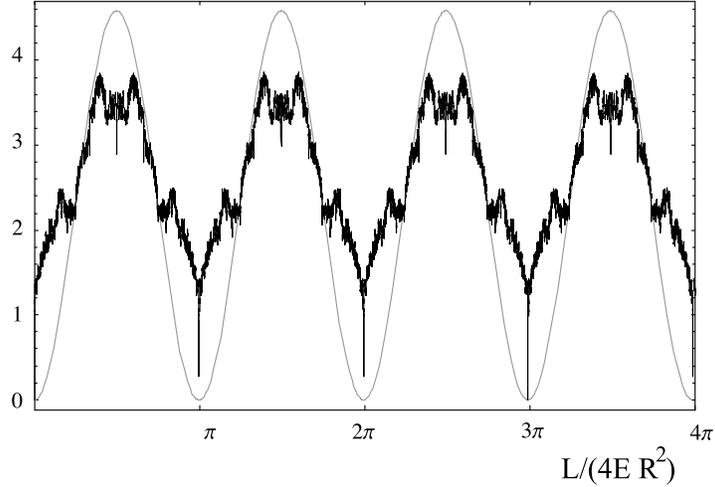}
\caption{\label{figf}{\it
Plot of the infinite two-dimensional sum
$\sum_{\vec{n}\in\mathbb{Z}^2}\frac{\delta^{-\frac{2h_{\vec{n}}}{M^2
R^2}}}{h_{\vec{n}}}
\sin^2\left(\frac{L}{4 E}\frac{h_{\vec{n}}}{R^2}
\right)$ with $h_{\vec{n}}=\cos^{-2}\theta(n_1^2+n_2^2-2 n_1 n_2
\sin\theta)$ in the case $M^2 R^2=10^6$, $\delta=16$ and
$\sin\theta=\frac{119}{120}$. The gray line represents the dominant
frequency mode $\rho \sin^2\left(\frac{L}{4 E}\frac{240}{239 R^2}\right)$
with amplitude $\rho={0.8 \pi}\cos\theta\log (M^2 R^2\cos\theta\log\delta^2)$.}}
\end{figure}

Assuming $\Delta_0\ll1$, we can drop the terms $F_2^2$ and $F_1^2<F_2^2$ in
(\ref{tsq}). Putting together (\ref{np}), (\ref{fsum}) and
(\ref{aproxr}), we obtain an approximate expression for the survival
probability:
\ba
P_{\nu_i\to\nu_i}\approx1-4\,{u_i^2(1-u_i^2)}{(1-\Delta_0)}\sin^2\frac{m_0^2 L}{4 E}
-3.2\,u_i^2\Delta_0\,\sin^2\left(\frac{\omega_{pq}}{R^2}\frac{L}{4 E}\right)
\label{surva}\, ,
\ea
where
\ba
m_0^2={m_D^2}{(1-\Delta_0)}\, .
\ea
Moreover, for the parameter $\zeta^2$ that enters in the average
transition probability formulas (\ref{cnuinui}), (\ref{cnuinuj}),
(\ref{pster}), we have
\ba
\zeta^2\sim\frac{1}{N_0^4}=1-2\Delta_0\ +\dots \ .
\label{zexp}
\ea

\end{document}